\documentclass[lettersize,journal]{IEEEtran}

\usepackage{amsmath,amsfonts}
\usepackage{array}
\usepackage{textcomp}
\usepackage{stfloats}
\usepackage{braket}
\usepackage{url}
\usepackage{verbatim}
\usepackage{graphicx}
\usepackage{mathtools}
\usepackage{amsmath,amsfonts}
\usepackage{array}
\usepackage{multirow}
\usepackage{array, caption}
\usepackage{graphicx}
\usepackage{makecell}
\usepackage{caption}
\usepackage{booktabs}
\usepackage{amssymb, nccmath}
\usepackage{textcomp}
\usepackage{url}
\usepackage{subfigure}
\usepackage{pgfplots}

\usepackage{pgfplots}
\usepackage{graphicx}
\usepackage{textcomp}
\usepackage[absolute,overlay]{textpos}
\usepackage{comment}
\DeclareUnicodeCharacter{FB01}{fi}
\usepackage{graphicx}
\usepackage{caption}
\usepackage{csquotes}
\usepackage{colortbl}
\usetikzlibrary{arrows,shadows} 
\usepackage{tikz}
\usetikzlibrary {positioning}
\usetikzlibrary{trees}
\usetikzlibrary{arrows}
\usepackage{longtable}
\usepackage{caption}
\usepackage{amsmath,amsfonts}
\usetikzlibrary{patterns}
\usepackage{pgfplots}
\usepackage{xcolor}
\usepackage{multicol}
\usepackage{multirow}
\usepackage{graphicx}
\usepackage{amsmath}
\usepackage{amssymb}
\definecolor{mypink}{HTML}{ADFF2F}
\definecolor{mycolor1}{HTML}{F5F5DC}
\definecolor{Mycolor}{HTML}{8A2BE2}
\usepackage{cuted}
\usepackage{longtable}
\definecolor{myhund}{HTML}{BE0032}
\definecolor{myfifty}{HTML}{FF003F}
\definecolor{mycolor1}{HTML}{F5F5DC}
\definecolor{mytwenty}{HTML}{8B008B}
\definecolor{myzero}{HTML}{FF007F}
\definecolor{pssfhun}{HTML}{0000FF}
\definecolor{pssffif}{HTML}{333399}
\definecolor{pssftwen}{HTML}{8A2BE2}
\definecolor{pssfzero}{HTML}{6699CC}
\definecolor{ff}{HTML}{D2691E}
\definecolor{bf}{HTML}{FFBF00}
\definecolor{rf}{HTML}{967117}
\definecolor{s1}{HTML}{CE2029}
\definecolor{s2}{HTML}{006D5B}
\definecolor{s3}{HTML}{8DB600}
\definecolor{r4}{HTML}{CE2029}
\usepackage{graphicx}
\usepackage{cite}
\usepackage{amsthm}
\usepackage{amssymb}
\usepackage[linesnumbered, ruled]{algorithm2e}
\makeatletter
\newcommand{\removelatexerror}{\let\@latex@error\@gobble}
\makeatother
\usepackage{dsfont}

\begin{document}
	\title{Secure Resource Management in Cloud Computing: Challenges, Strategies and Meta-Analysis}


		\author{Deepika Saxena, \IEEEmembership{ Member, IEEE}, Smruti Rekha Swain, Jatinder Kumar, Sakshi Patni, Kishu Gupta, \IEEEmembership{Member, IEEE}, Ashutosh Kumar Singh, \IEEEmembership{Senior Member, IEEE}, and Volker Lindenstruth, \IEEEmembership{Senior Member, IEEE}
			\thanks{Received 24 September 2024; accepted 31 December 2024. This work was supported in part by the JSPS Early Career and Young Researcher Grant 2024 under Grant H-2024-13, and in part by the Competitive Research Funding 2024 provided by the University of Aizu, Japan. This article was recommended by Associate Editor Y. Wan. (Corresponding author: Sakshi Patni.)}
			\thanks{Deepika Saxena is with the Division of Information Systems, University of Aizu, Aizuwakamatsu 9650006, Japan and also with Department of Computer Science, the University of Economics and Human Sciences, 01-043 Warsaw, Poland (e-mail: deepika@u-aizu.ac.jp).}
			\thanks{Smruti Rekha Swain and Jatinder Kumar are with the Department of Computer Applications, National Institute of Technology Kurukshetra, Kurukshetra 136119, India (e-mail: smruti.sai90@gmail.com and jatinderkumar2851@gmail.com).}
			\thanks{Sakshi Patni is with the Department of Computing, Gachon University, Seongnam 461701, South Korea (e-mail: sakshichhabra555@gmail.com).} 
	 \thanks{Kishu Gupta is with the Department of Computer Science and Engineering, National Sun Yat-sen University, Kaohsiung 804, Taiwan (e-mail: kishuguptares@mail.nsysu.edu.tw).}
\thanks{Ashutosh Kumar Singh is with the Department of Computer Science and Engineering, Indian Institute of Information Technology Bhopal, Bhopal 462003, India, and also with Department of Computer Science, the University of Economics and Human Sciences, 01-043 Warsaw, Poland (e-mail: ashutosh@iiitbhopal.ac.in).}
  \thanks{Volker Lindenstruth is with the Department of Computer Science and Engineering, Goethe University, 60438 Frankfurt, Germany (e-mail: voli@compeng.de). }
			\thanks{Color versions of one or more figures in this article are available at
				https://doi.org/10.1109/TSMC.2025.3525956.\\
				Digital Object Identifier 10.1109/TSMC.2025.3525956}
		}

 \markboth{IEEE TRANSACTIONS ON SYSTEMS, MAN, AND CYBERNETICS: SYSTEMS}%
{Shell \MakeLowercase{\textit{SAXENA et al.}}: SRM IN CLOUD COMPUTING: CHALLENGES, STRATEGIES AND META-ANALYSIS}


\makeatletter
\def\ps@IEEEtitlepagestyle{%
	\def\@oddfoot{\mycopyrightnotice}%
	\def\@oddhead{\hbox{}\@IEEEheaderstyle\leftmark\hfil\thepage}\relax
	\def\@evenhead{\@IEEEheaderstyle\thepage\hfil\leftmark\hbox{}}\relax
	\def\@evenfoot{}%
}

\def\mycopyrightnotice{%
	\begin{minipage}{\textwidth}
		\centering \scriptsize
		This article has been accepted for inclusion in a future issue of IEEE TRANSACTIONS ON SYSTEMS, MAN, AND CYBERNETICS: SYSTEMS Journal. Content is final as presented, with the exception of pagination. \\
		2168-2216 © 2025 IEEE. All rights reserved, including rights for text and data mining, and training of artificial intelligence and similar technologies. Personal use is permitted, but republication/redistribution requires IEEE permission. \\
		See https://www.ieee.org/publications/rights/index.html for more information.
	\end{minipage}
}
\makeatother
		
		\maketitle
		
		\begin{abstract}
			Secure resource management (SRM) within a cloud computing environment is a critical yet infrequently studied research topic. This paper provides a comprehensive survey and comparative performance evaluation of potential cyber threat countermeasure strategies that address security challenges during cloud workload execution and resource management. Cybersecurity is explored specifically in the context of cloud resource management, with an emphasis on identifying the associated challenges. The cyber threat countermeasure methods are categorized into three classes: \textit{defensive strategies}, \textit{mitigating strategies}, and \textit{hybrid strategies}. The existing countermeasure strategies belonging to each class  are thoroughly discussed and compared. In addition to conceptual and theoretical analysis, the leading countermeasure strategies within these categories are implemented on a common platform and examined using two real-world virtual machine (VM) data traces. Based on this comprehensive study and performance evaluation, the paper discusses the trade-offs among these countermeasure strategies and their utility, providing imperative concluding remarks on the holistic study of cloud cyber threat countermeasures and secure resource management. Furthermore, the study suggests future methodologies that could effectively address the emerging challenges of secure cloud resource management.
  
		\end{abstract}
		
		
		\section{Introduction}	\IEEEPARstart{C}{loud} 
		Computing has become an indispensable platform for all the industries, academia, business enterprises, social web-networks, etc. to serve the precise needs and provide the service features, tools and applications as per their requirements to help them meet their workloads processing, networking, and storage demands \cite{szefer2011eliminating,zhang2022two, saxena2022high, bi2024arima,swain2024intelligent,yadav2020adaptive}.  Within cloud data centers (CDC)s, the workload or online transactions are processed at outsourced cloud resources equipped with virtualization technology enabling resource sharing among multiple users \cite{feng2018privacy,{yadav2021managing}, shen2017block,liu2015two}. The virtualization of cloud resources  improve resource utilization and enables users to  switch their demands dynamically and pay accordingly \cite{saxena2023performance,barroso2013datacenter, chhabra2023secure,ding2018dfa, yu2014security}. Otherhand, the discrepancies and unpatched susceptibilities developed during virtualization prompts security breaches, sensitive data hampering, and severe cyber threats. The vulnerable VMs and hypervisors expedite the occurrence of cyberattacks \cite{saxena2023ai,saxena2023emerging}. An adversary or hacker may initiate multiple VMs for exploiting the VMs and hypervisor vulnerabilities in multiple ways to launch a cyber threat on one or multiple target VM(s) \cite{yin2022cloud,yuan2020minimizing,zhang2022trust}. The mismanagement during
physical resource distribution yields co-residency and creation of side-channels resulting into adversary VM appealing security threats such as leakage of user's confidential data, hampering of
data, unauthorized access via insecure interfaces, hijacking of 
accounts, etc. \cite{zhao2024cooperative, saxena2024high, yu2014security, liang2017mitigating}.  Therefore, procuring the SRM within cloud infrastructure has become a key challenge for  the stakeholders and Cloud Service Providers (CSP)s.

\subsection{Motivation}
Cyber threats are gobbling up the utility of the cloud services for the beneficiaries, including cloud service providers  as well as the end users. According to the estimation of Norton Security, in 2023, cyber criminals will
be breaching 33 billion records per year \cite{cloud2021security}. A recent survey focusing on security breaches of cloud data marked the biggest cloud cyber threats of millions of users' sensitive data that have surpassed on-premise breaches \cite{attacks2022}. It reports that even the giant cloud service providers  including Facebook, Alibaba, LinkedIn, Accenture, Sina Weibo, etc.  have failed to serve the secure service management to the users. These cyber threats have witnessed  the misconfiguration and mismanagement
associated with the virtualization technology at the cloud platform are the topmost causes of leakage of terabytes of sensitive data of millions of cloud users across the world \cite{cloud2020security,yadav2018adaptive, alouffi2021systematic, yadav2017mereg}. 
The security loopholes that generate during inefficient workload distribution and resource management  expedites VM security threats. Correspondingly, if the mismanagement and inefficiency during resource allocation are deliberately handled by detecting underlying  vulnerability with the concern of security, ample of security breaches that hamper users' sensitive information can be interrupted or halted. These facts generate a strong motivation for a  comprehensive survey analysis of existing countermeasure strategies against security threats  to address the the critical  objectives of imparting security during physical resource allocation in cloud environments.

\subsection{Secure Cloud Resource Management Viewpoint }
\begin{figure}[!htbp]
	\centering
	\includegraphics[width=0.99\linewidth]{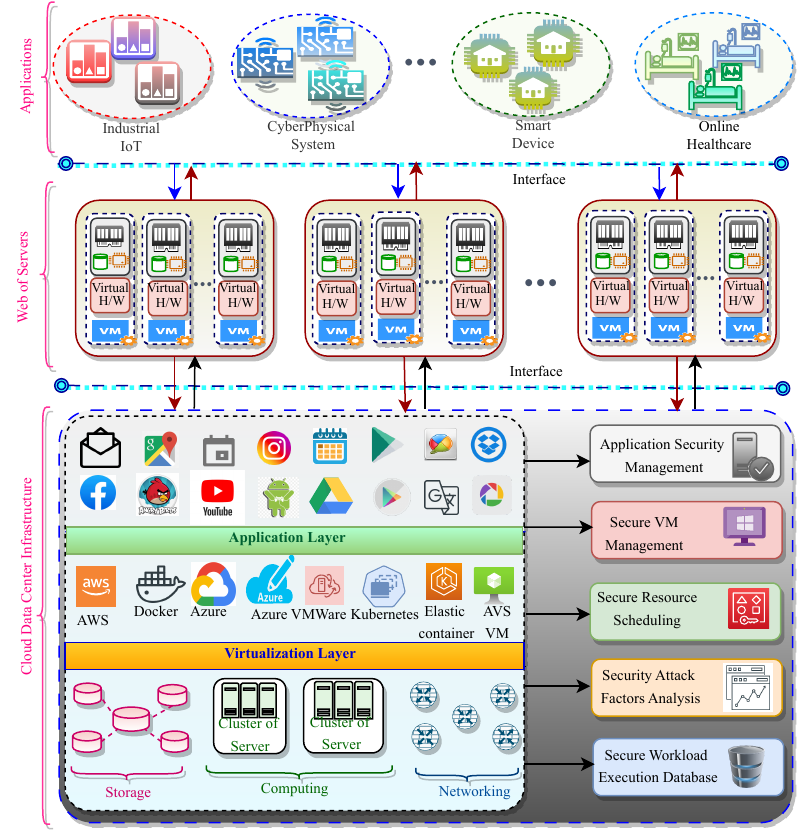}
	\caption{Security perspective in cloud resource management}
	\label{fig:fig1securityoverview}
\end{figure}

The context and utility of securing cloud resource management is exhibited in Fig.  \ref{fig:fig1securityoverview} through an interactive information flow among users' \textit{A}pplications, \textit{W}eb of servers, and \textit{C}loud data center infrastructure. 
The heterogeneous workload applications from a wide range of sources such as industrial Internet of Things (IoT), cyberphysical systems, smart devices, and online healthcare, etc. send their data processing requests to the web of servers within cloud infrastructure. Each server is equipped with high computing and storage resources, the hypervisor enables deployment of various number and capacities of VMs for execution of user applications {\cite{yadav2017mums}}. The cloud infrastructure comprising of scalable storage, networking, and computing resources serving execution of extensive range of cloud applications such as GMail, Youtube, Facebook, etc. is supported by the virtualization layer deployment using Docker, Amazon Web Services (AWS), Kubernetes, etc.  Considering a broader view of scheduling and distribution of multiple users' applications, it has been observed that the virtual resources are allocated negligently or lacking security perspective of  workload processing.  However, as discussed in the motivation that the security loopholes that are generated during inefficient workload distribution and resource management open the doors to VM security threats. To this context, security becomes the major concern during resource allocation and mangement. As illustrated in Fig. \ref{fig:fig1securityoverview}, the intended security is addressed at various level of service management with the help of \textit{Application Security Management}, \textit{Secure VM Management}, \textit{Secure Resource Scheduling}, \textit{Security Attack Factors Analysis}, and \textit{Secure Workload Execution Database}. By inducing and utilizing various levels of security management during cloud resource management, the cybersecurity associated with sensitive workload processing and confidential transactions get enhanced.

\subsection{Research Challenges}
The cloud computing environment is highly obliged to provide secure workload processing and data communication to serve the reliable and secure services to the users. To maintain effective security along with performance-efficiency is a compelling feature that drives the users' trust and dependence with reliability for the heterogeneous cloud services. Nevertheless, securing the cloud infrastructure and intended resource management is a complex problem entangled with some major challenges as discussed below:
\begin{itemize}
    \item \textit{Misconfigured Resources}: {Misconfiguration of physical and virtual resources is one of the most significant security challenges in cloud environments. Such misconfigurations can lead to vulnerabilities that adversaries exploit. For instance, the Capital One breach in 2019 resulted from a misconfigured web application firewall, allowing attackers to access sensitive data of over 100 million customers \cite{capitalone2019}. There is a need of developing automated configuration management tools using machine learning algorithms to continuously monitor and rectify configuration vulnerabilities in real-time.}

    \item \textit{Heterogeneous/Unknown Users}: The  shared nature of the cloud complicates resource allocation among a diverse user base, including both legitimate users and potential adversaries. {For example, resource schedulers may struggle to identify malicious VMs from unknown users, leading to potential threats. The attacks against cloud services by the Lizard Squad illustrate how attackers can leverage shared resources to mask their activities \cite{lizard2014}. This challenge motivates to enhance user profiling and anomaly detection techniques using AI to accurately identify and mitigate malicious activities in heterogeneous environments.}

    \item \textit{External Data Sharing}: Weak security measures can facilitate unauthorized access to confidential data. The Equifax breach in 2017 is a stark reminder, where attackers exploited a vulnerability in an open-source web application framework to access sensitive personal data of 147 million consumers \cite{equifax2017}. {To effectively protect sensitive data during transfers, researchers must explore robust data-sharing frameworks that incorporate end-to-end encryption and fine-grained access controls.}

    \item \textit{Different Levels of Data Sensitivity}: The variation in data sensitivity across different users complicates the application of uniform security mechanisms. For instance, healthcare data is subject to strict regulations like HIPAA, while other data types may not require such stringent measures \cite{mishra2024global}.  {This challenge raises critical concerns for investigating adaptive security solutions that dynamically adjust protections based on the sensitivity of the data being processed can enhance overall security effectiveness.}

    \item \textit{Uncontrolled Network Connectivity}: Adversaries can exploit network pathways to launch attacks against virtual machines (VMs). {The Mirai botnet attack in 2016 demonstrated how compromised IoT devices could be used to exploit network connections and launch Distributed Denial of Service (DDoS) attacks on cloud services \cite{yuan2020minimizing}.} 

    \item \textit{Insecure Load Balancing}: Load balancers often prioritize efficiency, which can inadvertently introduce security vulnerabilities. For example, traditional load balancing strategies may overlook security checks during VM allocation, exposing the infrastructure to threats. {The solution countermeasure should focus on integrating security assessments into load-balancing algorithms to ensure optimal resource distribution while maintaining security, using techniques like threat modeling and risk assessment.}
\end{itemize}

 \subsection{SRM  Objectives} 
 The major research objectives of secure resource allocation and management in cloud environments include:

 \begin{itemize}
     \item \textit{Defense against malicious links among VMs}: To implement measures to detect and block malignant or suspicious connections between VMs to prevent the spread of malware and unauthorized data access.

     \item \textit{Preventing unauthorized access}: To enforce strict network segmentation and access controls to ensure that only authorized VMs can communicate with each other, for reducing the risk of unauthorized data transfers.

     \item \textit{Mitigating malicious effects in multi-tenancy}: To deploy isolation techniques and monitoring to minimize the impact of malicious activities in multi-tenant environments.

     \item \textit{Eliminating security vulnerabilities}: To conduct regular security assessments, manage security loopholes, and perform vulnerability scanning to identify and eliminate weaknesses in the cloud infrastructure.

     \item \textit{Minimizing the sharing of resources}: To employ resource allocation policies that limit the extent of shared resources among tenants,  the risk of cross-tenant attacks and data breaches gets reduced.
 \end{itemize}

\subsection{Contributions}
To the best of the authors' knowledge, this is the first paper which aims to carry out a comprehensive comparative study and associated performance evaluation of the cloud cyber threats countermeasure strategies within scope of the resource management. The key contributions of this paper are fourfold:

\begin{itemize}
	\item A novel classification of cloud cyber threat countermeasure approaches concerning SRM  is proposed and clearly depicted with illustrations. The commendable secure cloud resource management approaches are identified and grouped according to the developed counter-action strategy concepts.

	\item A critical discussion and comparison considering all the essential details of state-of-the-art works   are provided and their features are analyzed to determine the future research scope addressing the limitation of the respective class based SRM  approaches.
\item {The leading models in each category are implemented on a uniform platform to enable detailed experimental analysis and comparison of key performance metrics.
}

 \item The research provides insights into trade-offs, notable findings, open issues, and concludes with specific future research avenues concerning the SRM  techniques in cloud and intended emerging computing environments.	
	
\end{itemize}

\subsection{Paper Organization}
Section I discusses the cloud computing environment, cybersecurity perspectives, and challenges of securing resource management, using schematic representations and descriptions. It outlines research challenges, SRM  objectives, and key contributions. Section II discusses the research methodology, SRM  classification, and taxonomy, reviewing prominent literature on cloud cyber threat countermeasures. Section III covers defensive strategies, while Sections IV and V review mitigating and hybrid strategies, respectively. Section VI evaluates leading  SRM  models across three categories by implementing them at common platform, comparing their performance using key performance indicators (KPIs). Section VII discusses trade-offs among the intended three  SRM  strategies, emerging research challenges, and potential solutions. Section VIII identifies open issues and future research directions. Section IX concludes with remarks on secure cloud resource management.

 \section{Methodology and SRM  Classification}

	 This survey provides a thorough examination of cybersecurity approaches developed to address the wide range of security threats against cloud resource configuration, allocation, and management. A comprehensive range of secure cloud resource management strategies, models, and methodologies published in top-tier journals and conference databases such as IEEE, ACM Digital Library, Elsevier, Springer, and Wiley are explored and analyzed for comparative study. The research papers are organized based on their underlying cyber threat handling strategies and are broadly categorized into three distinct classes: \textit{Defensive}, \textit{Mitigating}, and \textit{Hybrid strategies}. 

  \begin{itemize}
      \item \textit{Defensive strategies}: The defensive SRM strategies that focus on preventing security breaches before they occur. These strategies aim to create a robust security posture that makes it difficult for attackers to penetrate the cloud infrastructure. These strategies emphasises  the implementation of strict access controls, comprehensive encryption protocols, advanced network security measures, well-defined security policies, and proactive vulnerability management. By fortifying the cloud environment, these measures ensure a resilient defense against potential threats.
      \item \textit{Mitigating strategies}: The strategies aim to minimize the impact of security breaches when they occur are referred as `mitigating strategies'. These strategies focus on quickly detecting, responding to, and recovering from security incidents to reduce damage and restore normal operations. They include incident response, backup and recovery, continuous monitoring and logging, vulnerability management, and comprehensive security training.

      \item \textit{Hybrid strategies}: The hybrid strategies combine both preventive and responsive approaches for comprehensive security. They employ an integrated security framework, proactive detection of VM threats and potential malicious client VMs, adaptive security measures using AI, collaboration and threat intelligence sharing, continuous improvement based on past incidents and ongoing assessments, and holistic security policies covering all aspects of secure cloud resource management.
  \end{itemize}
  
\begin{figure*}[!htbp]
		
		\centering
		
		\subfigure[Defensive Strategies ]{\includegraphics[width=.32\textwidth]{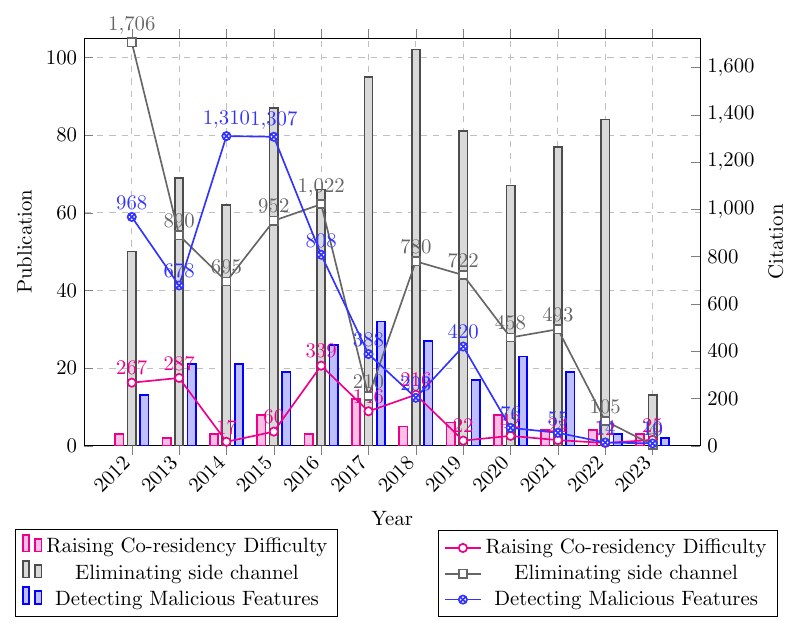}} 	\label{fig:defs} \hfill
		\subfigure[Mitigating Strategies]{\includegraphics[width=.32\textwidth]{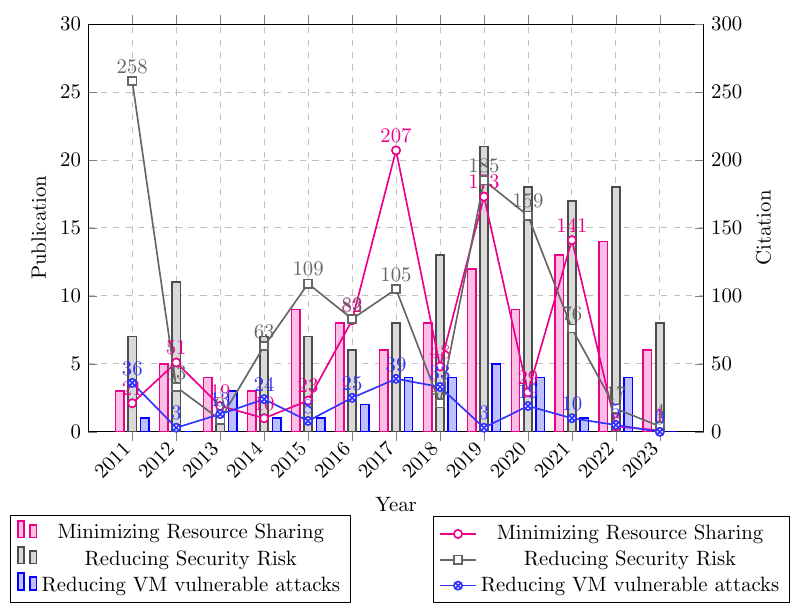}} \label{fig:mits} \hfill
		\subfigure[Hybrid Strategies]{\includegraphics[width=.32\textwidth]{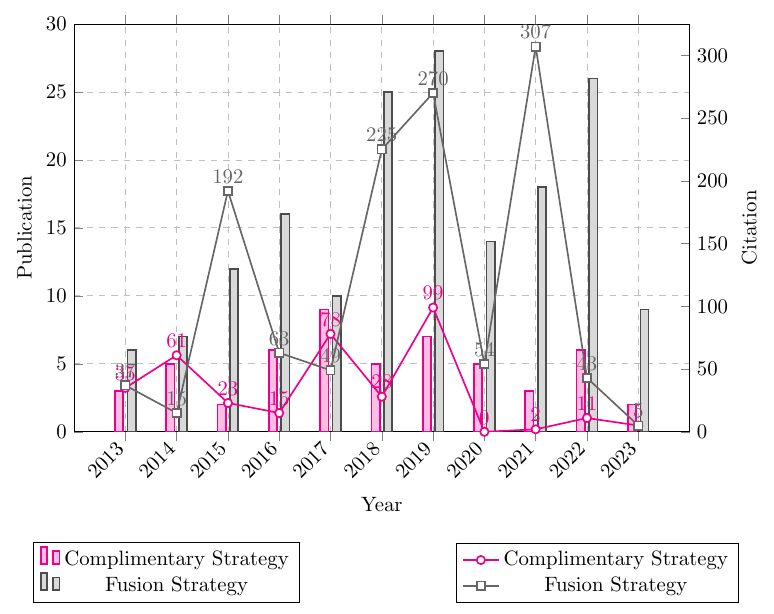}} \label{fig:hybs}\hfill
		
		\caption{Publications and citations over the timeline}
		\label{method}
		
	\end{figure*}

  The survey categorizes research papers on cloud cybersecurity into above three main classes based on similar approaches or overlapping features. Using the Scopus database, detailed bibliometric studies were conducted to gather information on publications and citations from 2002 to the present. Fig. \ref{method}(a), Fig. \ref{method}(b), and Fig. \ref{method}(c) present the bibliometric analysis for \textit{Defensive strategies}, \textit{Mitigating strategies}, and \textit{Hybrid strategies}, respectively. Specifically, the defensive strategies are reviewed in three divisions: \textit{Raising Co-residency Difficulty}; \textit{Eliminating Side Channel}; and \textit{Detecting Malicious Features}. The mitigating strategies are categorized into \textit{Minimizing Resource Sharing}; \textit{Reducing Security Risk}; and \textit{Reducing VM vulnerable attacks}. the hybrid strategies are distributed into \textit{Complimentary Strategy} and \textit{Fusion Strategy}. In the early 2000s, research on cloud security vulnerabilities and resource allocation was limited. However, significant attention and research on cyber threat countermeasures emerged in the second decade, leading to new methodologies for secure workload allocation and execution. This survey meticulously selects and investigates key works for a comprehensive literature review and meta-analysis. The increasing number of publications and citations across all three strategies reflects their growing importance and popularity.

 \section{Cyber Threats Defensive  Strategies}
Defensive security strategy focuses on deflecting and detecting the security loopholes and system susceptibilities that leads to sensitive data breaches and loss of data integrity. The  defensive security aims to defend against known attack vectors and prevent or restrict the occurrence of any data breach. This is achieved by using threat intelligence, vulnerability management, advanced testing, network segmentation,  security awareness, etc. The subsequent sections presents  illustrative depiction of the defensive strategy and its significant contributions.
\subsection{Depiction}
\begin{figure}[!htbp]
	\centering
	\includegraphics[width=0.999\linewidth]{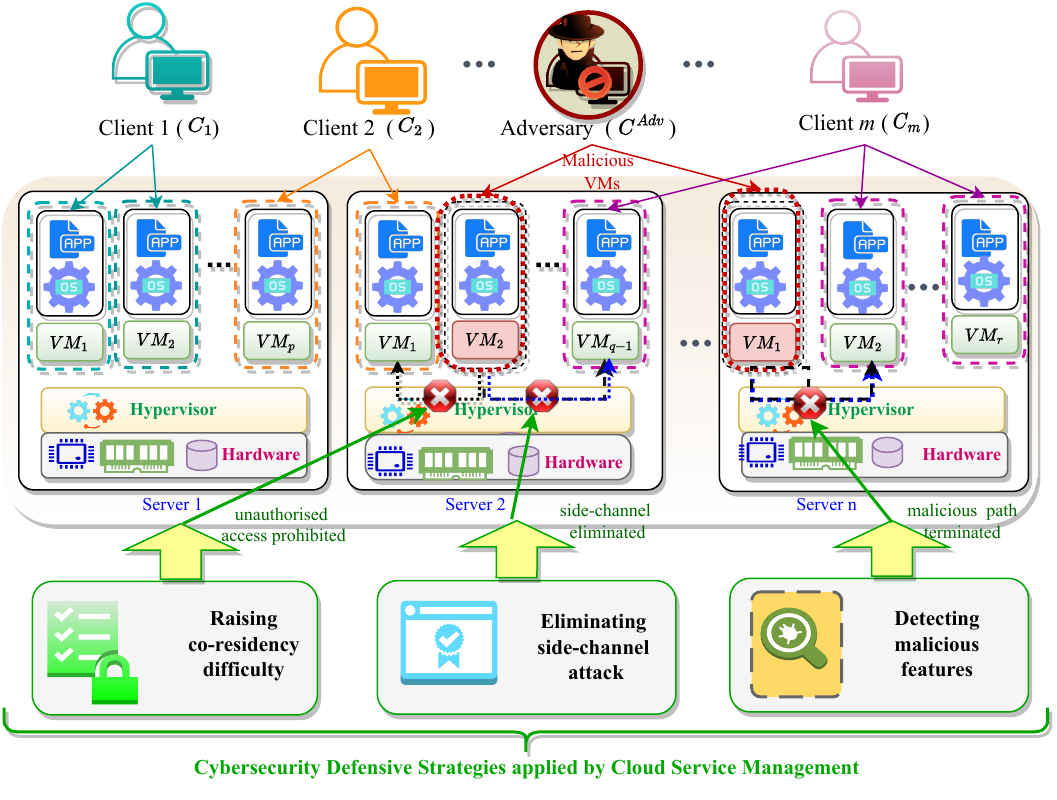}
	\caption{Defensive strategies against cyber threats}
	\label{fig:ds}
\end{figure}

Consider  $m$ clients \{$C_1$, $C_2$, ..., $C_m$\} operates various VMs such as \{$VM_1$, $VM_2$, ..., $VM_q$\} which are hosted on {$n$} servers.  Among the clients, assume  $C^{Adv}$ is a \textit{adversary} who  launches multiple applications on VMs, thus making them malicious ($VM^{Mal}$).  These malicious VMs compromise the vulnerable benign VMs ($VM^{Ben}$) by achieving co-residency and exploiting VM and hypervisor vulnerabilities as shown in Fig. \ref{fig:ds}. Correspondingly, the Cloud Resource Manager  (CRM) employs defensive security strategies to prevent the  data breaches via security threats launched by the adversary.
 As illustrated in Fig. \ref{fig:ds}, the cloud service management unit encapsulates key defensive countermeasures against cyber threats, including \textit{raising co-residency difficulties}, \textit{eliminating side channels}, and \textit{detecting malicious features}. {The mathematical formulation for detecting malicious co-residency and related features is presented in Eqs. (\ref{e1}-\ref{e3}). Eq. \eqref{e1} represents the VM allocation, denoted as $\omega_{ij}^k$, where the $i^{\text{th}}$ VM ($VM_i$) of the $k^{\text{th}}$ client ($C_k$) is hosted on the $j^{\text{th}}$ server ($S_j$). In Eq. \eqref{e2}, the term $\uplus_{ij,k \rightarrow i^{\ast}j,k^{\ast}}$ specifies the relationship between the VMs of the $k^{\text{th}}$ client and the $k^{\ast^{\text{th}}}$ client. This relation takes values from the set $\{ \text{non-malicious} (0), \text{malicious} (1) \}$. The status of the relationship between the two co-resident VMs, $i^{\text{th}}$ and $i^{\ast^{\text{th}}}$, hosted on the same server $S_j$, is denoted by $\Theta_{i,i^{\ast}}$. Eq. \eqref{e3} computes the cyber threat status, $\mathds{CT}$, such that if $\Theta_{i,i^{\ast}} = 1$ (indicating a malicious relationship between $i^{\text{th}}$ and $i^{\ast^{\text{th}}}$ VMs) and $k^{\ast}$ belongs to an adversary client ($C^{\text{Adv}}$), then a cyber threat is successfully identified. Otherwise, no cyber threat is detected. Based on the computed cyber threat status, $\mathds{CT}$, appropriate defensive measures are taken.} If a threat is detected, the system responds by raising difficulties for the intended co-residency and eliminating side channels to defend against the security threat.
 The related literature is discussed in the following subsections.

\begin{gather}
\omega_{ij}^k \mapsto VM_i \times S_j \times C_k \label{e1} \\
\Theta_{i,i^{\ast}} = \omega_{ij}^k \times \omega_{i^{\ast}j}^{k^{\ast}} \times \uplus_{ij,k\rightarrow i^{\ast}j,k^{\ast}} \label{e2}\\
\mathds{CT}= 	\begin{cases}
1, & {If \quad k^{\ast} \in C^{Adv} || \Theta_{i,i^{\ast}}==1 }\\
0, & { {\textit{Otherwise}} } 
\end{cases} \label{e3}
\end{gather}

\subsection{Key contributions}

 Han et al. \cite{han2017using} introduced the Previously-Selected-Servers-First (PSSF) scheme to counter co-residency driven VM attacks by limiting benign VMs' exposure to malicious VMs. Servers maintain a record of users whose VMs they have previously hosted, prioritizing these servers for future VMs of the same users. Similarly, Agarwal et al. \cite{agarwal2019secure} proposed the Previously Co-Located Users First (PCUF) method, where VMs are placed based on previous user allocations to reduce co-residency attacks. Yu et al. \cite{yu2013detecting} proposed the VMs Co-residency Detection Scheme (VCDS) to identify VM locations using cache-based side channel attacks and calculate co-residency probabilities. Han et al. \cite{han2016game} developed a two-player Game-based Defense Mechanism (GDM-VMP) for VM placement against side-channel attacks, using clustering and semi-supervised learning to differentiate attackers from legitimate users. Yuchi et al. \cite{yuchi2015enabling} quantified cloud security risks from VM vulnerabilities and proposed a placement algorithm minimizing these risks based on the US National Vulnerability Database (NVD). Natu et al. \cite{natu2017secure} suggested incorporating security requirements into VM placement policies, proposing Multiple VM Placement Security (MVPS) techniques using Linear Programming, Bin Packing, Graph Coloring, and Genetic Algorithm to reduce co-residency attacks.

Shi et al. \cite{shi2011limiting} presented Chameleon, a dynamic page coloring mechanism for eliminating cache-based side-channel attacks during security-critical operations. Vattikonda et al. \cite{vattikonda2011eliminating} proposed eliminating fine-grained timers (EFGT) by making the timer provided by the RDTSC instruction coarser. Wang and Lee \cite{wang2006covert} suggested Selective Partitioning and Random Permutation Cache (RPCache) to prevent multi-threading covert channels and thwart cache-based side channel attacks. Han et al. \cite{han2020quantify} introduced a deep learning model, Quantifying Co-residency Risks-based Security (QCS), to combat co-located VM attacks by profiling client behavior patterns. Wu et al. \cite{wu2012xenpump} developed XenPump to mitigate timing channel threats by adding latency to hypercalls. Kong et al. \cite{kong2019secure} proposed the Secure Container Deployment Strategy (SecCDS) using Genetic Algorithm to reduce co-residency in container clouds. Varadarajan et al. \cite{varadarajan2014scheduler} proposed a scheduler offering minimum runtime guarantees to prevent cache-based side-channel attacks. The comprehensive and summarised comparison of the aforementioned countermeasure strategies are elucidated in Table \ref{summaryDS}.


\begin{table*} [!htbp]
    
\caption{Summary and comparison of cyber threats defensive strategies}
		\label{summaryDS}
\scriptsize
\scriptsize
\resizebox{0.999\textwidth}{!}{
	\begin{tabular}{p{1.6cm} p{1.3cm}  p{3.5cm} p{1.7cm}p{3.0cm} p{2.3cm}p{1.520cm} p{3.0cm}} 
		
			\\ \toprule
		{\textbf{Model} }&	\textbf{Attack dealt}&{\textbf{Countermeasure strategy}}&\textbf{Datasets}&\textbf{Implementation}&{\textbf{Performance metrics}}& \textbf{{Limitations}}&{\textbf{Conclusive Remarks}} \\ \hline \hline
		PSSF (Han et al.	\cite{han2017using}, 2017) & VM Co-residency attack & Users' VMs are allocated to previously assigned servers first, otherwise, they are hosted on servers which consume least energy& VMs traces from
		NeCTAR & CloudSim, OpenStack&  Attack coverage, power consumption, attack efficiency& {static policy and non-adaptive} &PSSF furnishes an efficient VM placement policy with reduced co-residency attack and energy-efficiency  \\ \hline
		
		PCUF	(Agarwal et al.	\cite{agarwal2019secure}, 2019) & VM Co-residency attack & Probability of malicious VM co-residency is reduced by appointing previously co-located VMs to be placed on common server &  Microsoft Azure VM usage traces&CloudSim&Resource usage, co-location resistence& {static policy and non-adaptive}  &  Co-residency based attacks get reduced by co-location resistance with little compromise in resource utilization\\ \hline
		
		GDM-VMP	(Han et al.		 \cite{han2016game}, 2016) & Malicious Co-residency VM attack& clustering and semi-supervised learning techniques for client classification& NeCTAR data traces & CloudSim& Attack coverage rate& {complex implementation and management} & Attacker’s difficulty increases dramatically by one-to-two
		times of the magnitude of their defence mechanism\\ \hline
		VCDS (Yu et. al \cite{yu2013detecting}, 2013) & Cache-based side channel attack &  Load predictors based on linear regression &  JMeter was used to simulate multi-users to 
		access the web site   &Cloud 
		computing platform based on KVM, which consists of 30 
		servers & Co-residency probability \& computation time & {complex implementation and resource-intensive}& VCDS improved  true detection rate of the co-resident noisy VM compared with the 
		existing schemes \\ \hline

	NVD+CVSS (Yuchi et al. \cite{yuchi2015enabling}, 2015) & VM vulnerability-based attack&VM vulnerabilities effect are minimized by adjusting VM placement &Random datatraces \& NVD repository & Details not mentioned& survivability possibility& {complex implementation and non-adaptive}& Can be extended by  minimizing energy consumption and network cost while respecting all the security constraints \\ \hline
	
	MVPS ( Natu et al. \cite{natu2017secure}, 2017)&Co-residency attack & Multiple VM placement constrained with security to minimize the potential of co-residency threat  &Six traces  from separate commercial private clouds & Details not mentioned&Loss of Efficiency \& Core Utilization& {non-reliable policy} & It is preliminary stage work and it needs to be implemented and verified further \\ \hline
		
		Chameleon	(Shi et.al \cite{shi2011limiting}, 2011) & Cache-based side-channel attack & Dynamic page coloring to partition caches among security critical applications and excluding cross-VMs cache interference & Synthetic data traces &Xen running hardware-assisted VMs (HVM) with shadow page management  &Time of transferring files& {complex implementation and resource-intensive} & This work can be extended to investigate the tradeoff between the numbers of dedicated
			cache colors and the incurred overhead \\ \hline
			
	EFGT	(Vattikonda et. al \cite{vattikonda2011eliminating}, 2011) &Eliminated Fine Grained Timers 	& Have modified the RDTSC instruction on Xen-virtualized x86 machines&  Real-world workloads using micro- and macro-benchmarks & Testbed of two machines having 2.5GHz quad-core processor with 6GB memory running Linux version 2.6.32 and Xen 4.0.1, and connected via 1 Gbps Linksys switch  &Throughput and Round trip time& {complex implementation and resource-intensive} & This work  eliminated fine grained timers in Xen virtualized machine to forbid side-channels \\ \hline
			
		 Random Permutation Cache \cite{wang2006covert} (Wang et al., 2006) & Cache-based covert channel attacks	& Developed RPCache 
		 to tackle cache-based  side 
		 channel  and selective partitioning  for SMT-based 
		 covert channel  attacks&  SPEC2000  datatraces & SimpleScalar simulator uses a 2-way set associative write-back L1 cache with data marked by RPCache&Normalized execution time \& security analysis& {complex and moderate performance} &By analyzing the information leakage issues at the processor architecture level, two novel solutions to handle covert and side-channels are implemented \\ \hline
		
	 QCS	(Han et al. \cite{han2020quantify}, 2020) & Co-Resident Attacks & Deep Neural Network and DBSCAN algorithm for clustering of subscribers& Azure Public Dataset &Dell Precision Tower T5810 Workstation with an Intel Xeon E5-1620, 32G RAM, and Nvidia Quadro P5000 Graphic Card for GPU & Precision, recall, F-score, cross-entropy, accuracy& {complex and resource-intensive} & The method works only when all the subsribers are previously known and it is not adaptable as per the new subscribers are added to the group \\ \hline
	 	
		XenPump (Wu et al. \cite{wu2012xenpump}, 2012) & Timing channels & XenPump is presented to mitigate the threat of timing channel by adding latency and limit  user privacy leakage	& Random data traces &  Xen hypervisor with kernel 2.6.34.1 and 512 MB is running in each VM  & Latency & {complex and moderate performance}& XenPump prevents timing channel-based threat by adding latencies to the Hypercalls without collecting users' privacy data \\ \hline
		
		SecCDC (Kong et al. \cite{kong2019secure}, 2019) & Co-residency attack & SecCDS is proposed using
		Genetic Algorithm to defend against co-resident attacks &  Random data traces & Set of VMs, users, containers, and security requirements of cloud datacenters & Co-resident Rate \& migration times & {complex and data-dependent}&  Co-residency of cloud is  reduced with security level adjustment and minor impact on workload and performance,
		and is scalable for larger clouds \\ \hline
		Minimum Run
	Time (MRT) 	(Varadarajan et. al \cite{varadarajan2014scheduler}, 2014) & Cross-VM attack & Minimum run
	time  guarantee scheme is proposed for VM, virtual CPUs that limits the frequency of preemptions can  prevent Prime+Probe  side-channel attacks & CPU-hungry Workloads and Latency-sensitive Workloads & Testbed of Xen 4.2.1, Intel Xeon E5645, 2.40GHz clock, 6 cores in one package with
  32 KB L1 , 256 KB  L2, 12 MB shared L3 and 16 GB main memory & Latency sensitivity \& batch efficiency& {complex and non-durable performance} & It provides a high performance of achieving soft isolation, which limits the frequency of potentially dangerous cross-VM interactions \\ \hline
  \end{tabular}}
	\end{table*}
\normalsize

\section{Cyber Threats Mitigating  Strategies}
The cybersecurity threat mitigation strategies focus on minimizing the overall security risk and mitigating vulnerabilities leading to data breaches through the implementation of enhanced security policies and procedures. The detailed explanations and their respective key contributions are presented in the following subsections.
\subsection{Depiction}
\begin{figure}[!htbp]
	\centering
	\includegraphics[width=1.0\linewidth]{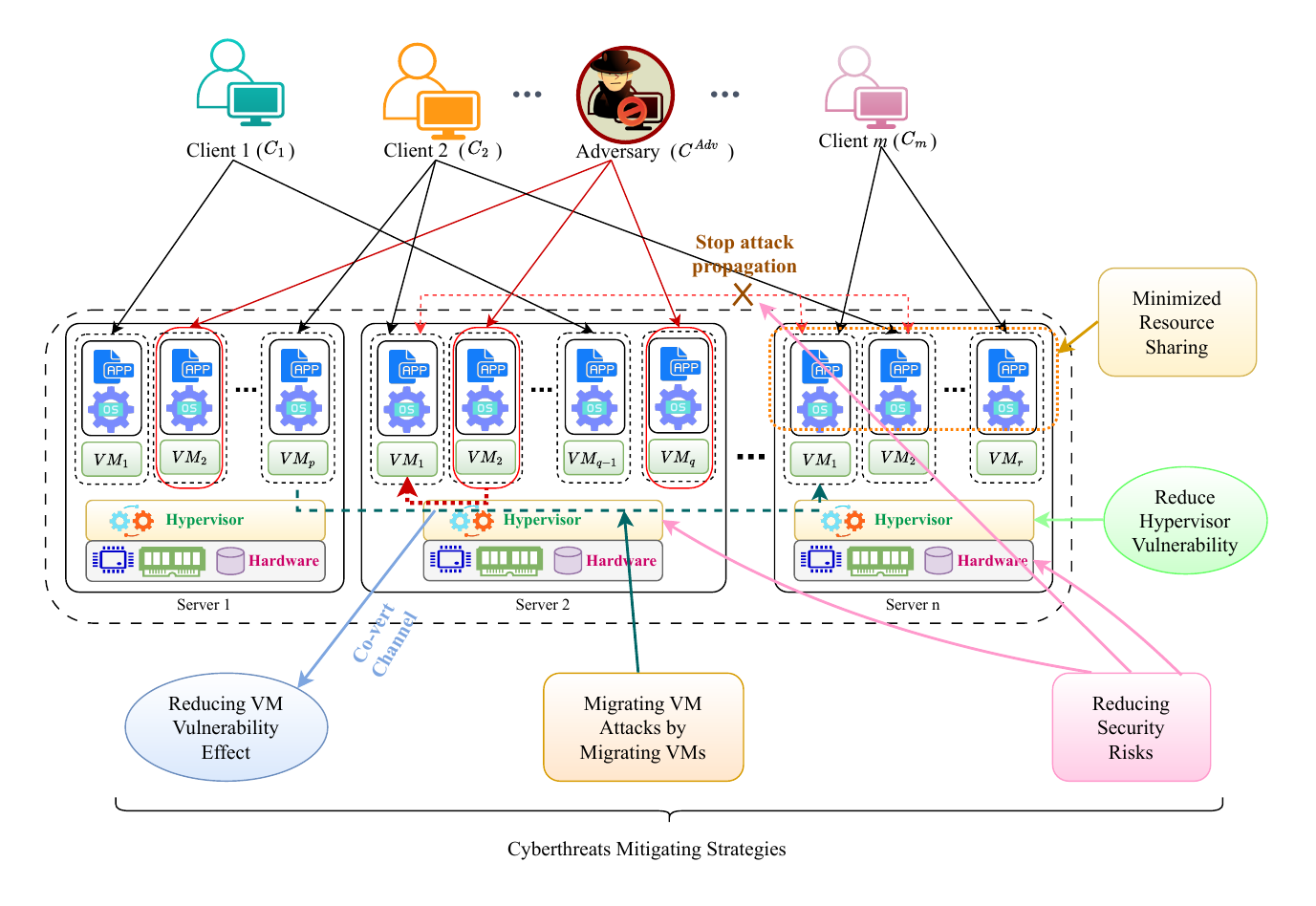}
	\caption{Mitigating strategies against cyber threats}
	\label{fig:ms}
\end{figure}

As illustrated in Fig. \ref{fig:ms}, $m$ clients \{$C_1$, $C_2$, ..., $C_m$\} launches different number of  VMs such as \{$VM_1$, $VM_2$, ..., $VM_q$\} which are accompained by {$n$} servers.  Let  $C^{Adv}$ {be} an \textit{adversary} who has launched multiple malicious VMs ($VM^{Mal}$) which establish unauthorised links to  compromise the co-resident benign VMs ($VM^{Ben}$)  and exploit VM and hypervisor vulnerabilities as shown in Fig. \ref{fig:ms}. Correspondingly, the Cloud Resource Manager  (CRM) employs  cybersecurity threat mitigation strategies to prevent the  data breaches via security threats launched by the adversary. These stratigies aim to reduce VM  and hypervisor vulnerabilities effects. The cyber threats can be addressed by enabling \textit{minimum resource sharing}, \textit{reducing security risks}, \textit{mitigating VM attacks}.

\subsection{Key contributions}
Several secure VM placement strategies have been proposed to mitigate security threats in cloud environments. Saxena et al. \cite{saxena2021secure} introduced the Secure and Multiple Objectives-based VM Placement (SVMP) framework, reducing co-residency and side-channel attacks. Singh et al. \cite{singh2019secure} presented the Secure and Energy-aware Load Balancing (SEA-LB) strategy, optimizing power saving and resource utilization with a non-dominated sorting genetic algorithm. Zhang et al. \cite{zhang2020secure} proposed the Load Balancing Model for Computation Offloading and Security (LBCOS), integrating encryption and load balancing to reduce response time and energy demands. Chhabra et al. \cite{chhabra2023secure} developed the Secure and Energy Efficient Dynamic Hierarchical Load Balancing (SEE-DHLB) framework, enhancing security through multi-objective load balancing. Arvindhan et al. \cite{arvindhan2022analysis} introduced the Hidden Markov State Transition-based Security (HMSTS) model for intrusion detection and prevention. Elshabka et al. \cite{elshabka2021security} proposed the Security-aware Dynamic VM Consolidation (SDVMC) technique for security risk management. Li et al. \cite{li2012improving} presented the Quantified Security Risks-based VM Placement (QSR-VMP), quantifying security risks to improve VM survivability. Han et al. \cite{han2017reducing} developed the Security-aware Multi-Objective Optimization based VM Placement algorithm (SMOOP) to optimize multiple security objectives. Zhao et al. \cite{zhao2017feature} introduced the Feature-based Transfer Learning-enabled Detection Framework for Network Security (FTL-NS) to detect unseen attack variants using feature-based transfer learning. The comprehensive and summarised comparison of the aforementioned research works are depited in Table \ref{summaryMS}.
\begin{table*}[!htbp]
		\caption{Summary and comparison of cyber threats mitigating strategies}
		\label{summaryMS}
\scriptsize
\resizebox{0.99\textwidth}{!}{
	\begin{tabular}{p{1.6cm} p{1.3cm}  p{3.5cm} p{1.7cm}p{3.0cm} p{2.3cm}p{1.520cm} p{3.0cm}} 
		
			\\ \toprule
		{\textbf{Model} }&	\textbf{Attack dealt}&{\textbf{Countermeasure strategy}}&\textbf{Datasets}&\textbf{Implementation}&{\textbf{Performance metrics}}& \textbf{{Limitations}}&{\textbf{Conclusive Remarks}}
\\ \hline \hline
		SVMP (Saxena et al. \cite{saxena2021secure}, 2021) & Co-residency attack & Multi-objective VM allocation is proposed with minimized resource sharing & Google datatraces & CDC environment is set-up within Python with benchmark server and VM configuration & Security attacks, resource utilization, power consumption& {complex \& resource-intensive} & Unknown malicious VM attack can be prevented \\ \hline
		
		SEA-LB (Singh et al. \cite{singh2019secure}, 2019) & Conflicting servers-based attack & Reduced resource sharing governed NSGA-II approach-based multi-objective load balancing & Random dataset & CDC environment is set-up within MATLAB with benchmark server and VM configuration & Security attacks, resource utilization, power consumption& {static \& less reliable} & Co-residency-based VM attack can be prevented  \\ \hline

		LBCOS (Zhang et al. \cite{zhang2020secure}, 2020) & Cloud vulnerabilities & AES cryptographic algorithm infused with ECG signal-based encryption for overcoming the vulnerability & Random dataset &workstation 
		with an Intel Core i7-4770 CPU with 3.4 GHz frequency, 16 GB RAM  running Windows 10 and MATLAB simulator & security for varying data capacity& {complex \& data-dependent} & Reduced security vulnerabilities with decreased response time and energy demands \\ \hline
		
		SEE-DHLB (Chhabra et al. \cite{chhabra2023secure}, 2023) & Co-residency attack & Task classification with identification and elimination of suspicious VMs & Synthetic datatraces & CDC is set-up using CloudSim in Java-Eclipse IDE & Throughput, resource utilization, response time& {static \& resource-intensive} &  Unknown or new malicious VMs are difficult to identify \\ \hline
		
		HMSTS (Arvindhan et al. \cite{arvindhan2022analysis}, 2022)	 & Intrusion-based attacks & Hidden Markov model is applied for identification and prevention of intrusion & Synthetic datatraces & Not mentioned & Rate of intrusion detection&{ complex \& resource-intensive} & Helpful for intrusion detection and mitigation in cloud \\ \hline

			SDVMC (Elshabka et. al \cite{elshabka2021security}, 2010) & Risks-based VM attack &Local Regression (LR) for overloads detection and Minimum Migration Time (MMT) for VM selection and proposed a novel MRI with RITH VM placement algorithm & Planet Lab VM traces  & CloudSim simulation with 800 servers,  of type HP ProLiant ML110 G4 and  HP ProLiant ML110 G5 & Security measure, Energy measure, QoS measure & {complex \& static policy} & Can be extended with strongest security assessment model considering intrusion detection and  vulnerabilities \\ \hline

		QSR-VMP (Li et al. \cite{li2012improving}, 2012) & Network-cascading based VM attacks 	& Quantified Security Risks-based VM placement with Markov chain prediction for VM threats& Random dataset & 81 VMs are hosted on 10 node &  Comparison of survivability \& cost on migration& {data\& resource-intensive}  & service survivability enhancement is 74.28\% and the average improvement of survivability possibility is
		27.15\% \\ \hline
		
		FTL-NS (Zhao et al. \cite{zhao2017feature}, 2017) & Network-security attacks & Heterogeneous transfer learning via spectral transformation & NSL-KDD  network attack dataset& Details not mentioned & Attack estimation accuracy& {complex \& data-dependent} &  transfer
		learning approach improves  new network attacks estimation compared with baselines algorithms \\ \hline
		FSVMP (Wong et al. \cite{wong2018secure}, 2018) & Co-residency attack & Familarity based Best Fit Placement (FBBF) algorithm, and Familiarity-based Load Balancing (FBLB) algorithm is developed & Random datatraces & CloudSim tool &Energy Consumption \& ability of co-residence attack&{ complex \& less reliable} &  improve the model by addressing more security problems \\ \hline

		SMOOP (Han et al. \cite{han2017reducing}, 2017) & Multiple security risks-based attacks & Security risks are quantified and secure multi-objective VM placement is proposed & Synthetic and random datatraces & VMs and Server machines are simulated in Java & Scalability \& distribution of security risks level& {static \& less reliable} & Generates Pareto-optimal secure VM placement solution \\ \hline


	\end{tabular}}
\end{table*}

\normalsize
\section{Hybrid Strategies}
Hybrid countermeasure strategies incorporate integrated approaches employed during cloud resource allocation to combat cyber threats. These strategies fall into two categories: \textit{Complimentary approaches} and \textit{Fusion approaches}, as depicted in Fig. \ref{fig:hs}. Complimentary approaches combine proactive cyber threat detection with reactive threat handling or mitigating mechanisms. The descriptions of these techniques and their existing key contributions are discussed in the following subsections.
\subsection{Depiction}
\begin{figure}[!htbp]
	\centering
	\includegraphics[width=0.999\linewidth]{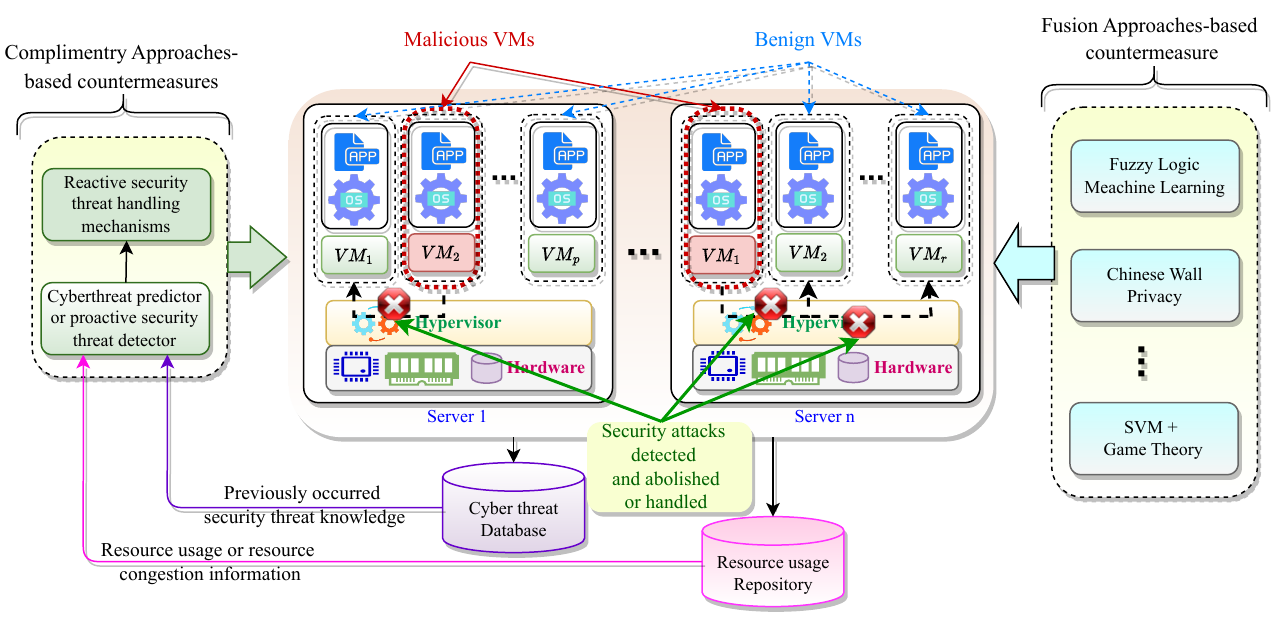}
	\caption{Hybrid strategies against cyber threats}
	\label{fig:hs}
\end{figure}

 As demonstrated in Fig. \ref{fig:hs}, complementary approaches involve exploring and investigating security risks and analyzing previously occurred threats to develop a knowledge database for proactively predicting cyber threats. Based on this analysis, appropriate threat-handling mechanisms are applied reactively. Additionally, workload execution or resource usage data is analyzed to predict and mitigate potential resource contention, which is often an indicator of a Denial-of-Service (DoS) type attack on virtual machines (VMs). The most commonly employed reactive method to address or mitigate security threats involves migrating vulnerable VMs to a secure host or server.

On the other hand, fusion approaches combine multiple techniques working collaboratively to achieve the shared goal of eliminating or mitigating cybersecurity threats (Fig. \ref{fig:hs}). The following sections provide a literature survey of existing complementary and fusion countermeasure approaches designed to address cloud cybersecurity threats.

\subsection{Key contributions}
Saxena et al. \cite{saxena2020security} devised a secure resource allocation scheme to address VM placement security concerns, incorporating a VM threat detector and workload predictor. They analyzed inter-VM relations and factors like malicious network traffic and bandwidth utilization for breach detection. Additionally, Saxena et al. \cite{saxena2023sustainable} proposed the Sustainable and Secure Load Management (SaS-LM) Model for enhanced client security and energy conservation in CDCs. Using Dual-Phase Black Hole Optimization, they optimized a neural network to predict congestion and adjust load while ensuring security and sustainability. In another work, Saxena et al. \cite{saxena2021oscmc} introduced the Online Secure Communication Model for Cloud (OSC-MC) to locate and terminate malicious VMs and inter-VM links, resulting in improved network performance and reduced energy consumption. Gill et al. \cite{gill2018secure} presented the SECURE mechanism, which detects and mitigates various threats in cloud environments using a detection engine and an autonomic system for attack response without user intervention.

Several fusion approaches have been proposed to address security challenges in cloud environments. Gaggero et al. \cite{gaggero2018model} introduced the Model Predictive Control-based VM Placement (MPC-VMP) approach, which proactively handles security and workload processing issues while reducing power requirements. Devi et al. \cite{devi2020deep} proposed a Deep Learning-based security model for task classification and scheduling, categorizing tasks into different security levels. He et al. \cite{he2017machine} suggested various machine learning algorithms for DDoS attack detection, leveraging statistical information from cloud servers and VMs. Kesavmoorthy et al. \cite{kesavamoorthy2019swarm} presented a Multi-Agent System based DDoS attack detection and prevention system (MAS-PSO) using Particle Swarm Optimization intelligence. Tariq et al. \cite{tariq2019agent} analyzed and mitigated multiple security risks using software and intelligent agents. Aldawood et al. \cite{aldawood2021secure} proposed a secure graph-based stacking algorithm (Gb-SRS) for secure VM allocation in cloud systems. Tao et al. \cite{tao2021secure} introduced a Secure and Efficient VM Allocation (SeVMA) mechanism to mitigate cross-VM attack threats while ensuring workload balance and energy consumption. Ahamed et al. \cite{ahamed2016security} suggested security-aware energy-efficient dynamic VM consolidation algorithms to reduce operating costs in CDCs. The comprehensive and summarised comparison of the aforementioned research works are elucidated in Table \ref{summaryHS}.
\begin{table*}
    
\caption{Summary and comparison of cyber threats hybrid strategies}
		\label{summaryHS}
\scriptsize
\resizebox{0.99\textwidth}{!}{
	\begin{tabular}{p{1.6cm} p{1.3cm}  p{3.5cm} p{1.7cm}p{3.0cm} p{2.3cm}p{1.520cm} p{3.0cm}} 
		
			\\ \toprule
		{\textbf{Model} }&	\textbf{Attack dealt}&{\textbf{Countermeasure strategy}}&\textbf{Datasets}&\textbf{Implementation}&{\textbf{Performance metrics}}& \textbf{{Limitations}}&{\textbf{Conclusive Remarks}}
 \\ \hline \hline
SEDRA (Saxena et al. \cite{saxena2020security}, 2020) & VM placement-based  threats & VM threats are predicted and mitigated before occurrence & Random datatraces & CDC is set up in CloudSim & Security attacks, resource utilization, power consumption& {complex \& resource-intensive} & VM threats are predicted and prevented proactively \\ \hline

SaS-LM (Saxena et al. \cite{saxena2023sustainable}, 2023) & Co-residency attack & Reduction of resource sharing and developing Multi-objective Dual-Phase  Blackhole VM allocation algorithm & Google datatraces & CDC environment is set-up within Python with benchmark server and VM configuration & Security attacks, resource utilization, power consumption & {complex \& resource-intensive}& Sustainability and security are concurrently considered during VM allocation \\ \hline

OSC-MC (Saxena et al. \cite{saxena2021oscmc}, 2023) & Coresidency \& Network-risk attack & Minimized the security threats by locating and terminating
malicious VMs and inter-VM links & Bitbrains VM datatraces & CDC environment is set-up within Python with benchmark server and VM configuration & Network hogs, unauthorized and authorized inter-VM links& {complex \& data-dependent} & Security is incorporated during VM inter-communication in cloud \\ \hline

SECURE	(Gill et al. \cite{gill2018secure}, 2018) & Multiple security attacks & A detection engine is employed to consistently detect security discrepancies while processing workloads & Datatraces of Image file of 713 MB & Server and VMs are implemented CloudSim & Intrusion detection rate  \& impact of security on response time& {data-dependent \& resource-intensive} & Multiple security attacks are dealt concurrently \\ \hline

MPC-VMP (Gaggero et al. \cite{gaggero2018model}, 2018) & Malicious VM attack & Predictive control approach for tackling hardware outages, rebooting of software, ensuring  security policies,  and reducing power requirements & Synthetic workload traces &Simulations
in MATLAB on a 2.5-GHz Intel Xeon PC with
16 GB of RAM & Number of secure VMs \& power consumption& {static \& resource-intensive} & MPC approach is portable to emerging
scenarios considering user mobility or geographically sparse infrastructures \\ \hline

DLSM (Devi et al. \cite{devi2020deep}, 2020) &Malicious VM attacks &Deep learning-based security model for tasks  classification and  scheduling & Synthetic workload & Cloud infrastructure in CloudSim 3.0 simulator & Respose time, resource utilization, \& execution time & {complex \& unreliable} & This work is preliminary stage work which needs further exploration \\ \hline

ML-DAD (He et al. \cite{he2017machine}, 2017)   & DDoS attack  & Various machine learning algorithms are used to detect DDoS attack &Synthetic datatraces &  Six servers and multiple  VMs in Openstack \& SSH brute-force, DNS reflection, ICMP flooding, and TCP SYN attacks are launched & Accuracy, Recall, \& F1 score of DDoS attack prediction & {data and resource-intensive} & DDoS attacks and traffic congestion are prevented \\ \hline

MAS-PSO (Kesavmoorthy et al. \cite{kesavamoorthy2019swarm}, 2019) & DDoS attack & Multi agent based DDoS attack detection
and prevention system using Particle swarm intelligence  & Synthetic workload requests  &Java based
Spring IDE with Cloudsim package & Attack detection accuracy \& time, \& false attacks& {complex \& resource-intensive} & MAS-PSO has the quickest attack detection as compared with existing methods\\ \hline

ABISF (Tariq et al. \cite{tariq2019agent}, 2019) & Multiple security risks & 
software and intelligent agents develop decision systems using security data& Random datatraces &Fuzzy inference system based 
upon fuzzy set theory \& logic rules in MATLAB & Risk identification, threat \& vulnerabilities determination agent & {complex \& resource-intensive}& Can be extended by adding an authentication, virtualization,  and privacy 
layers \\ \hline
Gb-SRS (Aldawood et al. \cite{aldawood2021secure}, 2021) & Side-channel attack &Graph-based security-aware heuristic, called Graph based Secure Random Stacking (GbSRS) and a migration algorithm called Graph-based VM Migration (GbM)&Azure VMs traces & Network CloudSim for network-related cloud simulations &Migration cost & {data \& resource-intensive}& a secure VM allocation algorithm with reduced cost of VMs migrations is developed \\ \hline

SeVMA (Tao et al. \cite{tao2021secure}, 2021) & Cross-VM attacks &Improved NSGA-II based VM placement \& suspicious VM detection-driven migration & Random data traces & CDC simulation with various number of VMs \& servers &Cross-VM attack threats, workload balance, \& energy consumption & {complex \& resource-intensive} & It is a preliminary stage work which needs to be improved with more security features \\ \hline

SBS (Ahamed et. al \cite{ahamed2016security}, 2016) & Malicious VM-based attack &  Consolidated ranking based security profile can be created for the VMs in the CDC that is used for security aware VM allocation &Planet Lab workload traces &CloudSim Simulator &Energy consumption, Energy times SLA violation& {complex \& data-intensive} & can be improved by including VM reliability as well as security and energy
consumption \\ \hline

\end{tabular}}

\end{table*}

\normalsize
\section{Performance Evaluation}

\subsection{Experimental set-up and datasets}\label{R1}
The simulation experiments are executed on a server machine assembled with two Intel\textsuperscript{\textregistered} Xeon\textsuperscript{\textregistered} Silver 4114 CPUs with 40 core processor and 2.20 GHz clock speed in Python. The computation machine is deployed with 64-bit Ubuntu 16.04 LTS, having main memory of 128 GB. The data centre environment is set up with three different types of servers and four types of VMs configuration shown in Tables \ref{table:server} and \ref{table:vm}. The resource features like power consumption ($P_{max}, P_{min}$), MIPS, RAM and memory are taken from real server IBM \cite{IBM1999} and Dell \cite{Dell1999} configuration where $S_1$ is 'ProLiantM110G5XEON3075', $S_2$ is 'IBMX3250Xeonx3480' and $S_3$ is 'IBM3550Xeonx5675'. The VMs configuration is inspired from the VM instances of Amazon website \cite{amazon1999EC2}. 

\begin{table}[!htbp]
	\centering
	
	\caption[Table caption text] {Server configuration}  
	\label{table:server}
	\resizebox{0.49\textwidth}{!}{
		\begin{tabular}{lccccccc}
			\hline
			Server&PE&MIPS&RAM(GB)&Memory(GB)&${PW}_{max}$&${PW}_{min}$/${PW}_{idle}$\\
			\hline
			$S_1$ 	& 2&2660&4&160&135&93.7 \\
			$S_2$	& 4&3067&8&250&113&42.3 \\
			$S_3$	& 12&3067&16&500&222&58.4 \\

			\hline
	\end{tabular}}
\end{table}

\begin{table}[!htbp]
	\centering
	
	\caption[Table caption text] {VM configuration}  
	\label{table:vm}
	\resizebox{0.49\textwidth}{!}{
	\begin{tabular}{lcccc}
		\hline
		VM type& PE &MIPS&RAM(GB)&Memory(GB)\\
		\hline
		${VM}_{small}$&1&500&0.5&40\\
		${VM}_{medium}$&2&1000&1&60\\
		${VM}_{large}$&3&1500&2&80\\
		${VM}_{Xlarge}$&4&2000&3&100\\

		\hline
	\end{tabular}}
\end{table}
{\textit{Datasets}: The selected approaches are evaluated using two benchmark VM traces from publicly available real workload datasets: the \textit{OpenNebula VM Profiling Dataset} (ONeb) \cite{24mb-vt61-20} and the \textit{Google Cluster Data} (GCD) \cite{reiss2011google}. The ONeb dataset contains information captured by a monitoring system via execution of VM probe programs provided by OpenNebula. It includes details on VM threats, server status, basic performance indicators, VM status, and capacity consumption of VMs and their hosting servers. While the exact values of VM and hypervisor vulnerability scores are not specified in the VM threat database, various risk scores for different VMs are computed using the method described by \cite{saxena2023ai}. This information is used by the VM threat predictor to estimate potential threats before they occur. The GCD dataset provides capacity usage information of CPU, memory, and disk I/O for 672,300 jobs executed on 12,500 servers over 29 days. The dataset includes CPU and memory utilization percentages recorded every five minutes for 24 hours. Users can request varying numbers and types of VMs over time, constrained by the total number of available VMs at the CDC. Experiments are conducted over 80 five-minute intervals to dynamically analyze the performance of various countermeasures. Since the original GCD traces do not include VM threat information, a VM threat database was generated for these traces, reporting attacks on VMs with attributes such as {${S}_id$, ${VM_id}$, ${VM}_i^{CPU}$, ${VM}_i^{BW}$, ${VM}_i^{memory}$, ${R}^{score}_i$, ${V}_i$, ${H}_i$, ${C}_i$, ${N}_i$, ${VM}_i^{status}$, etc}. VM vulnerability (${V}$) and server hypervisor vulnerability (${H}$) are represented as CVSS scores with a normal distribution in the range [0, 10]. Security threats depend on the values of the four types of risks and other relevant information in the threat database {${V}_i$, ${H}_i$, ${C}_i$, ${N}_i$} associated with a VM and its co-residency with malicious VMs (${VM}^{Mal}$)}.

\begin{itemize}
\item \textit{Resource Utilization}:
The resource utilization of data centre can be obtained using Eq. (\ref{ru1}) and  Eq. (\ref{ru2}), where if server (${S}_i$) is active, $\gamma_i=1$, else it is $0$ and $Z$ is number of resources and $\mathds{R} \in {{CPU}, {Mem}, etc}$. While the formulation considers only CPU (${CPU}$) and memory (${Mem}$), it is extendable to any number of resources.
\begin{gather}
{RU}_{CDC}= \int\limits_{\substack{t_1\\\mathcal{}}}^{t_2} (\frac{	{RU}_{CDC}^{{CPU}} +  {RU}_{CDC}^{{Mem}} }{|Z|\times \sum_{i=1}^{P}{\gamma_i}})dt\label{ru1}
\\
{RU}_{CDC}^{\mathds{R}}=\sum_{i=1}^{P}{\frac{\sum_{j=1}^{Q}{\omega_{ji} \times VM_j^{\mathds{R}}}}{S_i^{\mathds{R}}}} \quad  \label{ru2} 
\end{gather}
\item \textit{Power Consumption}:
From the existing literature \cite{sharma2016multi}, power consumption for $i^{th}$ server can be formulated as ${PW}_i$ and total power consumption ${PW}_{CDC}$ during time-interval \{$t_1$, $t_2$\} is shown in Eq. (\ref{power2}).
\begin{gather}
\resizebox{0.4\textwidth}{!}{$ 
{PW}_{CDC} = 
\int\limits_{\substack{t_1\\\mathcal{}}}^{t_2} (\sum_{i=1}^{P} {([{{PW}_i}^{max} - {{PW}_i}^{min}] \times {RU} + {{PW}_i}^{idle})}) dt
\label{power2} $}
\end{gather}	
where ${RU}$ is resource utilization, ${{PW}_i}^{max}$, ${{PW}_i}^{min}$ and ${{PW}_i}^{idle}$ are maximum, minimum and idle state power consumption for $i^{th}$ server.

\item \textit{Cyber threats coverage} ($\Xi$): Any malicious activity leading to data breaches or unauthorized access to data is termed a security threat. The likelihood of such threats varies based on the probability of multiple security risks and the learning capability of the threat detector, which depends on the number of cloud users and the size of CDCs. The total threat coverage ($\Xi_{coverage}^{CDC}$) over the time interval {$t_1$, $t_2$} is computed using Eqs. (\ref{attackcoverage1}) and (\ref{attackcoverage2}).

\begin{equation} \label{attackcoverage1}
\int_{t1}^{t2}	\Xi_{coverage}^{z'} = \int_{t1}^{t2}\frac{|CoresidTarget(VM_z, VM_{z'})|}{TotTarget(z')} 
\end{equation}
\begin{equation} \label{attackcoverage2}
\int_{t1}^{t2}	\Xi_{coverage}^{CDC}= \int_{t1}^{t2}  \Xi_{coverage}^{z'} 	
\end{equation}
where $Xi_{coverage}^{z'}$ is attack coverage of $z^{th}$ malicious user which is defined as the ratio of the number of co-residency of malicious user VMs ($ VM_{z'}$) with target VMs ($VM_z$) i.e $|CoresidTarget(VM_z, VM_{z'})|$ and total number of target VMs of $z'^{th}$ malicious user ($TotTarget(z')$) during time interval [$t_1$, $t_2$].	
\end{itemize}

\subsection{Comparative Results}
The performance of various cyber threat countermeasures across three major categories is thoroughly investigated and compared. These categories include \textit{Defensive strategies}: Previously Selected Server First (PSSF) \cite{han2015using} and Previously Co-located User First (PCUF) \cite{agarwal2019secure}; \textit{Mitigating strategies}: Security-aware Multi-Objective Optimization based VM Placement algorithm (SMOOP) \cite{han2017reducing} and Security and Energy Aware Load Balancing (SEA-LB) \cite{singh2019secure}; and \textit{Hybrid strategies}: Security Embedded Dynamic Resource Allocation (SEDRA) \cite{saxena2020security} and Secure and Sustainable Load Management (SaS-LM) \cite{saxena2023sustainable}. The investigation uses a wide range of heterogeneous cloud applications and varying VM resource utilization. We evaluated and compared these approaches based on cyber threat coverage over time, security success rate for different CDC sizes, resource utilization over time and across various CDC sizes, power consumption over time and for different CDC sizes, and the average number of active servers.

\subsubsection{Security} 
\begin{figure}[!htbp]	
 	\centering	
 	\subfigure[GCD traces]{\includegraphics[width=0.241\textwidth]{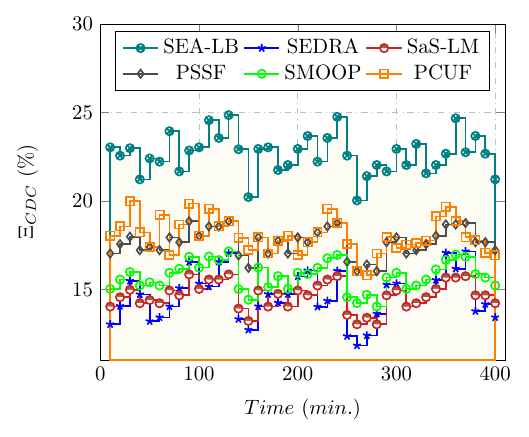}}
 	\subfigure[OpenNebula traces  ]{\includegraphics[width=0.241\textwidth]{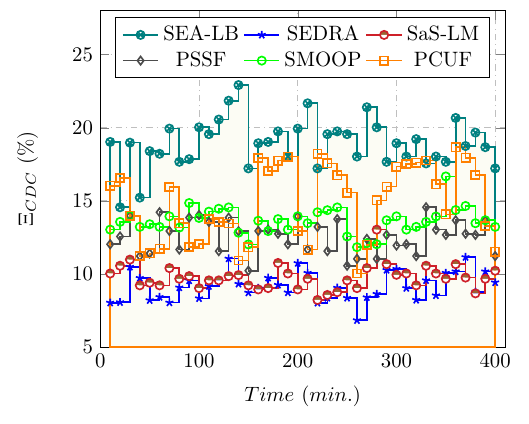}}\
 	\caption{Cyber threats v/s time }
 	\label{fig1}
 \end{figure}

\begin{figure}[!htbp]	
	\centering	
	\subfigure[GCD traces]{\includegraphics[width=0.241\textwidth]{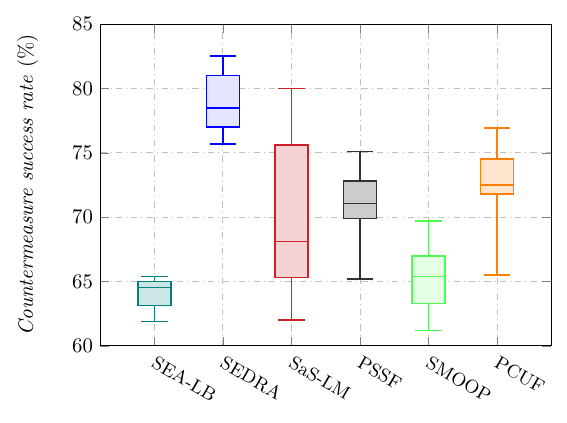}}
	\subfigure[OpenNebula traces]{\includegraphics[width=0.241\textwidth]{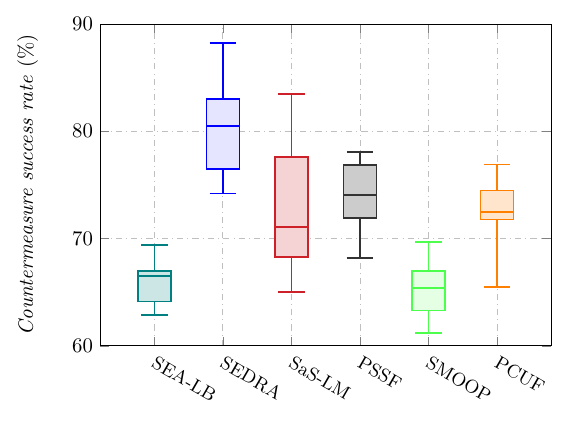}}
	\caption{Countermeasures success rate over size of CDCs}
	\label{fig2}
\end{figure}

The cyber threat coverage ($\Xi_{CDC}$ (\%)) of all considered countermeasure approaches is compared in Fig. \ref{fig1} for GCD traces (Fig. \ref{fig1}(a)) and ONeb traces (Fig. \ref{fig1}(b)). As observed in Fig. \ref{fig1}, $\Xi_{CDC}$ varies distinctly over time with different countermeasure methods for both traces. However, the hybrid strategies, SEDRA and SaS-LM, consistently produce fewer cyber threats ($\Xi_{CDC}$) compared to defensive and mitigating strategies for both data traces. The defensive strategies such as PSSF and PCUF surpass mitigating strategies including SEA-LB and SMOOP in terms of reducing $\Xi_{CDC}$ (\%). The cyber threat coverage ($\Xi_{CDC}$ (\%)) is reduced in the order: SEDRA $\geq$ SaS-LM $\geq$ PSSF $\geq$ PCUF $\geq$ SMOOP $\geq$ SEA-LB. Figs. \ref{fig2}(a) and \ref{fig2}(b) illustrate the comparative success rate of different countermeasure strategies against cyber threats for varying sizes of CDCs. The box plots represent a comprehensive comparison of the success rate against security attacks in terms of the bottom quartile, bottom whisker, median, top whisker, and top quartile. As discussed above, hybrid strategies like SEDRA and SaS-LM, which apply both proactive and reactive methods to combat upcoming threats and prohibit their successful launch, achieve the highest success rate for both datasets. The defensive strategies PSSF and PCUF achieve the next highest success rate, while the mitigating strategies SMOOP and SEA-LB show the lowest success rate.

The reason behind such trend outcomes is that hybrid strategies include proactive detection of threats and workload management based on previous knowledge of the threat database and workload information, which helps alleviate multiple-risk-based cyber threats and therefore reduce cyber threat coverage. Defensive strategies focus on minimizing the successful creation of malicious links and side channels, reducing co-residency-based attacks specifically. On the other hand, mitigating strategies such as SEA-LB attempt to minimize resource sharing, and SMOOP evaluates and mitigates vulnerabilities of VMs and hypervisors. These strategies aim to reduce the probability of the occurrence of cyber threats rather than eliminate them completely.


\subsubsection{Resource utilization}
\begin{figure}[!htbp]	
 	\centering	
 	\subfigure[GCD traces]{\includegraphics[width=0.24\textwidth]{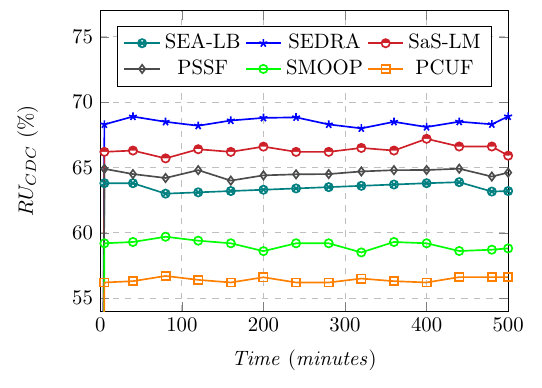}}
 	\subfigure[OpenNebula traces]{\includegraphics[width=0.24\textwidth]{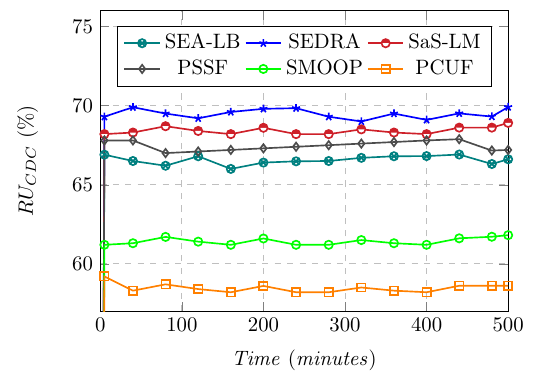}}
 	\caption{Resource utilization over time}
 	\label{fig3}
 \end{figure}

\begin{figure}[!htbp]	
	\centering	
	\subfigure[GCD traces]{\includegraphics[width=0.241\textwidth]{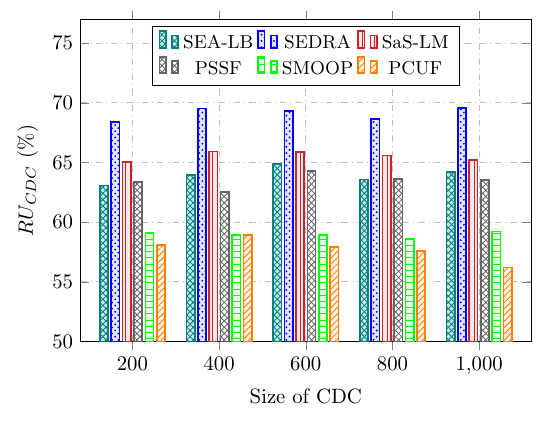}}
	\subfigure[OpenNebula traces]{\includegraphics[width=0.241\textwidth]{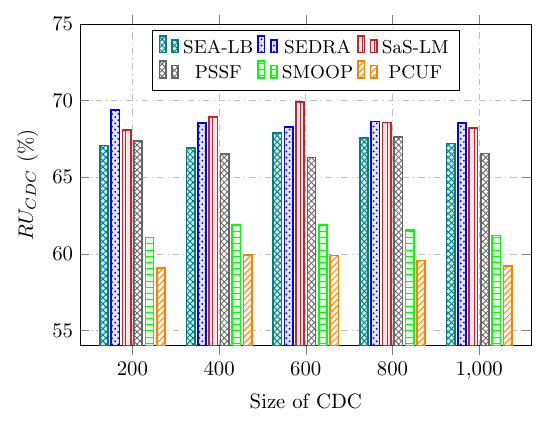}}
	\caption{Resource utilization over varying size of CDC }
	\label{fig4}
\end{figure}

Fig. \ref{fig3} compares the resource utilization ($RU_{CDC}$(\%)) over time for CDC of size 1000 VMs of the three categories of countermeasures using both GCD and ONeb traces. The resource utilization follows the trend: PCUF $<$ SMOOP $<$ SEA-LB $<$ PSSF $<$ SaS-LM $<$SEDRA for both data traces over time  which depicts $RU_{CDC}$(\%) is independent of  time-period.  Specifically, SEDRA  improves the average utilization of resources up to 12.67\%, 9.4\%, 5.3\%, 4.2\%, and 2.9\%, over PCUF, SMOOP, SEA-LB, PSSF, and SaS-LM, respectively using GCD traces. Similarly, SEA-LB surpasses the average resource utilization of PCUF, SMOOP, SEDRA, PSSF, and SaS-LM by 11.2\%, 8.4\%, 3.8\%, 2.5\%, and 1.3\%, respectively for ONeb data traces. Nonetheless, SEDRA exhibits highest security with 5.3\% and 3.8\% higher resource utilization than SEA-LB. Fig. \ref{fig4} depicts the comparison of resource utilization ($RU_{CDC}$ (\%)) with varying numbers of VMs or sizes of CDCs, reinforcing that resource utilization is independent of CDC size. The hybrid strategy SEDRA outperforms all other strategies in terms of average resource utilization for both data traces across a wide range of VMs within the CDC. Furthermore, in SEDRA and SaS-LM, physical resource allocation to VMs is based on predicted resource utilization, resulting in higher resource efficiency. In contrast, defensive and mitigating strategies like PCUF, SMOOP, PSSF, and SEA-LB do not apply workload prediction, deploying VMs with user-demanded resource capacities, which are often much greater than the actual resource utilization, leading to higher resource wastage.

\subsubsection{Power consumption}
\begin{figure}[!htbp]	
	\centering	
	\subfigure[GCD traces]{\includegraphics[width=0.241\textwidth]{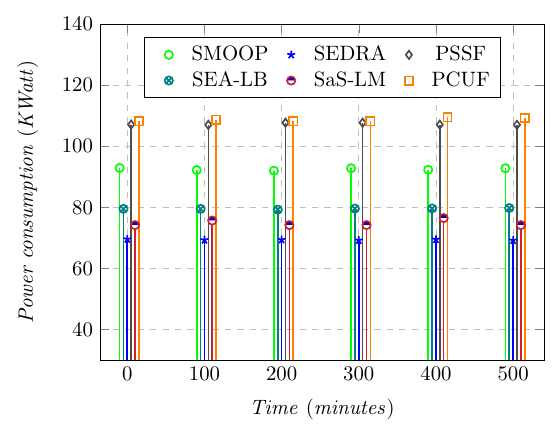}}
	\subfigure[OpenNebula traces]{\includegraphics[width=0.241\textwidth]{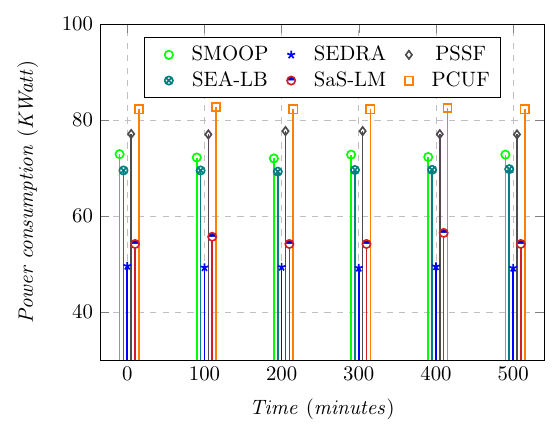}}	
	\caption{Power consumption over time}
	\label{fig5}
\end{figure}

\begin{figure}[!htbp]	
	\centering	
	\subfigure[GCD traces]{\includegraphics[width=0.241\textwidth]{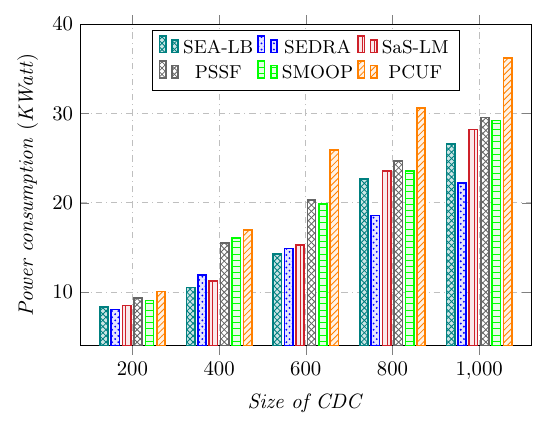}}
	\subfigure[OpenNebula traces]{\includegraphics[width=0.241\textwidth]{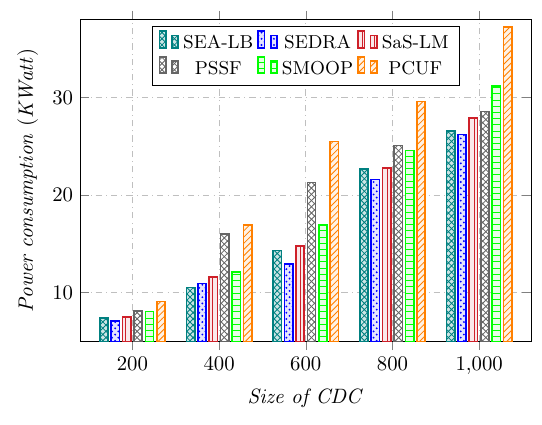}}
	\caption{Power consumption over varying size of CDC}
	\label{fig6}
\end{figure}

The periodic values of power consumption ($PW_{CDC}$ (KWatt)) of defensive, mitigation, and hybrid strategies are compared in Fig. \ref{fig5} for CDCs of size 1000 VMs via bar graphs, where hybrid strategies including SaS-LM and SEDRA have reduced power consumption over defensive strategies: PSSF and PCUF, and mitigating strategies: SMOOP and SEA-LB.  The $PW_{CDC}$ obtained for the varying size of CDC for the compared approaches   is reported in Fig. \ref{fig6} that depicts increase of $PW_{CDC}$ with increasing size of the CDC because of the increasing number of active servers to serve the intended workloads execution. The reason behind such a reduction in power consumption is the minimization of resource wastage and number of active servers in case of the
hybrid strategies due to resource utilization prediction and deployment of VMs with required physical resource capacity equivalent to predicted resource usage of each VM. The reduction in power consumption follows the trend similar to resource utilization: PCUF $<$ SMOOP $<$ SEA-LB $<$ PSSF $<$ SaS-LM $<$SEDRA. 

\subsubsection{Active servers}
\begin{figure}[!htbp]	
	\centering	
	\subfigure[GCD traces]{\includegraphics[width=0.241\textwidth]{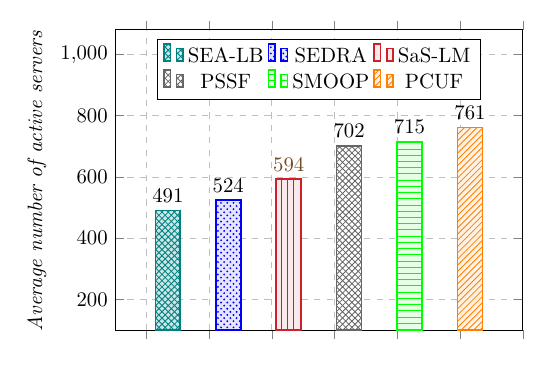}}
	\subfigure[OpenNebula traces]{\includegraphics[width=0.241\textwidth]{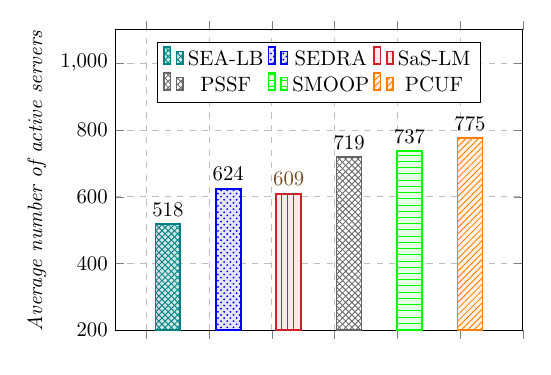}}
	\caption{Average number of active server machines}
	\label{fig7}
\end{figure}

The average number of active servers (${S}_{active}$) are compared in Fig. \ref{fig7}, wherein  SEA-LB and SEDRA show least while PCUF reveal highest number of active servers.  The number of active server follows the trend: PCUF $>$ SMOOP $>$ PSSF $>$ SaS-LM $>$ SEDRA $>$ SEA-LB.  SEDRA provides highest security among all the comparative methods at the cost of 6.72\% (in case of GCD) and  20.8\% (in case of ONeb) more active servers as compared with SEA-LB method. SaS-LM stands third in the list of average number of active servers and provide security equivalent to SEDRA.  

\section{Trade-offs and Insights}

In cloud environments, three primary countermeasure strategies including defensive, mitigating, and hybrid, present distinct trade-offs {\cite{yadav2020energy}} between security and resource efficiency. Each strategy offers a different approach to achieving secure resource management, with strengths and limitations based on their focus and adaptability to emerging threats.

\textit{Defensive strategies} prioritize preventing specific security risks, such as co-residency and hypervisor vulnerabilities. These strategies excel at stopping well-known attacks without requiring prior knowledge of threat patterns. However, their rigid nature limits adaptability to unforeseen or evolving threats. Additionally, defensive techniques like raising co-residency difficulty or eliminating side-channel attacks often increase power consumption and activate more servers, leading to low resource utilization and higher operational costs. While effective for single threat scenarios, they exhibit high time and space complexity due to the need for meticulous VM allocation planning.

\textit{Mitigating strategies} focus on reducing the attack surface by optimizing load distribution. For example, by assigning all VMs of a single user to a common server, the risk of inter-user attacks decreases. However, this comes at the cost of placing all VMs at risk if that server is compromised. Mitigating strategies offer better resource efficiency compared to defensive methods but can expose users to higher security risks if a common server is breached.

\textit{Hybrid strategies} combine the strengths of both defensive and mitigating approaches by leveraging proactive and reactive techniques to predict and address threats. By using historical threat data and machine learning models, such as transfer learning, hybrid methods anticipate new security risks while dynamically adjusting resource allocation to maintain efficiency. This balance allows for better optimization of power consumption, resource utilization, and VM consolidation. However, hybrid strategies depend on the accuracy of threat prediction, making them sensitive to false positives or inaccurate analysis. Their success in balancing security and efficiency hinges on the quality of threat detection algorithms and workload estimation techniques.

\subsection{{Trade-off Between Security and Performance}}

{The trade-off between security and performance is a critical consideration in the deployment of cloud security strategies. Defensive strategies enhance security by implementing stringent controls, but this often results in increased latency and resource consumption, affecting overall performance. For instance, rigorous VM placement to prevent co-residency may lead to sub-optimal resource utilization and higher operational costs. Conversely, mitigating strategies aim to optimize performance by improving resource allocation and minimizing user co-residency, which can inadvertently compromise security. By focusing on resource efficiency, these strategies might expose cloud environments to greater risks, especially if a shared server is compromised.} {Hybrid strategies strive to achieve an optimal balance between these two factors. By employing machine learning and predictive analytics, they enhance security while maintaining performance efficiency. However, the effectiveness of these approaches is contingent upon the accuracy of threat predictions and the system's ability to adapt to dynamic environments.}

\subsection{Key Takeaways and Insights}

Key insights from this analysis include:

\begin{itemize}
    \item \textit{Defensive Strategies}: Best suited for well-known, single-risk attacks. While they do not require prior threat knowledge, they are resource-intensive, leading to high power consumption and the need for many active servers.
    \item \textit{Mitigating Strategies}: Improve resource utilization by minimizing resource sharing but introduce the risk of significant data loss if a common server is compromised. These strategies reduce complexity but sacrifice some security in multi-tenant environments.
    \item \textit{Hybrid Strategies}: Provide a balanced, scalable solution by combining historical data with real-time threat detection. They enhance resource efficiency and security through machine learning and predictive analytics but require continuous monitoring and accurate threat models to be effective.
\end{itemize}

\subsection{{Achieving Trade-offs through Optimization}}

{The optimization process for these strategies involves balancing security requirements against resource constraints such as power consumption, active physical machines (APMs), and resource utilization (RU). Defensive strategies optimize for security by carefully allocating VMs to servers, reducing the risk of co-residency or side-channel attacks. This requires solving complex VM placement problems that maximize security while minimizing the number of active servers, thus controlling power consumption and ensuring optimal resource use. Mitigating strategies, conversely, optimize for resource efficiency by distributing VMs in a way that reduces multi-user co-residency. This approach typically minimizes the number of active servers but increases the risk of single-point failures if those servers are compromised. Hybrid strategies employ machine learning models, such as transfer learning, to predict potential threats and adjust VM allocation dynamically. These models optimize the allocation process by predicting threat patterns and consolidating workloads to reduce power consumption while maintaining security through continuous threat monitoring. The optimization challenge lies in balancing the accuracy of threat predictions with resource allocation decisions to ensure efficient use of infrastructure while minimizing vulnerabilities.}

The trade-offs across these strategies reflect a balance between achieving robust security and maintaining efficient resource management. Defensive strategies prioritize security but may compromise efficiency, mitigating strategies favor resource optimization but risk security lapses, and hybrid strategies seek an optimal balance by integrating predictive analytics and real-time adjustments. Future research should focus on refining these optimization processes to enhance both security and resource efficiency in cloud environments. The key parameter-based trade-off among  existing cyber threat countermeasure strategies is presented in Table \ref{tradeoff}.
\begin{table*}
		\caption[Table caption text] {{Key parameter-based trade-off among  existing cyber threat countermeasure strategies} }
			\label{tradeoff}
	\centering 	
	\resizebox{0.95\textwidth}{!}{
	\renewcommand{\arraystretch}{1.8}
	\begin{tabular}{|c|c|c|c|c|c|p{1.7cm}|p{1.7cm}|p{1.95cm}|p{1.8cm}|p{1.8cm}|}
		\hline
		\multicolumn{2}{|c|}{\textbf{Strategies}} & \textbf{RU} & \textbf{PW} & \textbf{APM} & \textbf{PTB}& \textbf{Time complexity}  & \textbf{Space complexity}  &  \textbf{Sophisticated  planning} & \textbf{Single risk-based threat}  & \textbf{Multiple  risks-based threat} \\
		\hline \hline
		\multirow{3}{*}{\rotatebox[origin=c]{90}{\shortstack{\textbf{Defensive}}}} & DS1  & $\downarrow$ &  $\uparrow$ &  $\uparrow$ & $\times$&high  & moderate  & very high   & same & different \\
		& DS2  & - &  $\uparrow$ &  $\uparrow$  & $\times$& moderate  & moderate  & very high   & same  & different \\
		& DS3  & - &  - &  -  &$\times$& high/ moderate  & high/ moderate  & high   & same  & different \\  \hline \hline
		\multirow{3}{*}{\rotatebox[origin=c]{90}{\shortstack{\textbf{Mitigating}}}} & MS1  & $\downarrow$ &  $\uparrow$ &  $\uparrow$  &$\times$& moderate  & moderate  & very high   & same  & different \\
		& MS2  & - &  - &  -  &$\times$& moderate  & moderate  & very high   & same  & different\\
		& MS3  & - &  - &  -  &$\times$& moderate  & moderate  & very high   & same  & different \\  \hline \hline
		\multirow{2}{*}{\rotatebox[origin=c]{90}{\shortstack{\textbf{Hybrid}}}} & HS1  & $\uparrow$ &  $\downarrow$ &  $\downarrow$  &$\checkmark$& very high  & very high  & high  & same  & same \\
		& HS2  & - &  - &  -  &-& very high  & very high  & high & same  & same or different \\
		\hline
	\end{tabular}}
	\\
\footnotesize{{{RU: Resource utilization, PW: Power consumption, APM: Active physical machines,$\uparrow$: High, $\downarrow$: Low, -: Not determined, DS1: Raising coresidency difficulty, DS2: Eliminating side-channel, DS3: Detecting malicious feature, MS1: Minimizing resource sharing, MS2: Reducing VM attacks, MS3: Mitigating VM attacks, HS1: Complimentary approach, HS2: Fusion approach, PTB: Prior threat database}}}	
\end{table*}
The mitigating strategy of minimizing resource sharing (MS1) aims to reduce physical machine sharing among users, leading to more active servers and higher power consumption. It does not require prior threat knowledge and has moderate time and space complexities. On the other hand, reducing VM attacks (MS2) and mitigating VM attacks (MS3) require sophisticated VM allocation planning for different security threats. Hybrid strategies (HS1 and HS2) predict threats and workloads for high resource use and low power consumption but involve high time and space complexities. Proactive threat estimation in hybrids needs prior threat knowledge. Despite complexity, hybrids outperform individual strategies in security, resource use, and power reduction.
The choice of a SRM  strategy should be based on the sensitivity of the workload and the acceptable cost of time and space complexity. This decision should align with the terms and obligations of secure workload execution agreed upon by cloud service providers and stakeholders.

\section{Open Issues and Future Directions}

As cloud computing continues to expand, resource management is becoming increasingly complex due to the need to balance performance, scalability, and security. Modern cloud infrastructures must support multi-tenant environments, high availability, and dynamic scaling—all while addressing emerging cyber threats. Securing cloud resource management has become one of the most critical challenges, as vulnerabilities in resource allocation can be exploited, leading to performance degradation, data breaches, and service outages.

This complexity is compounded by the rapid evolution of attack vectors, making static security measures insufficient. To address these concerns, research in secure cloud resource management is focusing on innovative methods that integrate dynamic threat adaptation, AI-driven defense mechanisms, and collaborative frameworks. This section highlights the key open issues and future research trends in this domain, with a focus on advancing cloud resilience and security while maintaining optimal resource utilization. {The key \textit{open issues and future directions} are as follows:}

\begin{itemize}
    \item {\textit{Balancing Security and Performance in Resource Allocation}: One of the foremost challenges in secure cloud resource management is creating algorithms that effectively balance security requirements with performance needs. As cloud workloads fluctuate, future research must focus on developing \textit{dynamic resource allocation algorithms} that can adjust in real-time based on both performance demands and evolving cyber threats. These algorithms should account for varying resource needs across different deployment environments, from small-scale setups to large, multi-tenant architectures.}

    \item {\textit{Scalable Security Solutions for Diverse Deployment Scenarios}:  Cloud environments are diverse and dynamic, ranging from public and private clouds to hybrid and multi-cloud setups. Future research must prioritize \textit{scalable security frameworks} that are flexible enough to protect cloud resources across different deployment scales and architectures. These solutions should be lightweight yet robust, ensuring that security protocols can be applied consistently without introducing significant overhead.}

    \item {\textit{Dynamic Threat Adaptation Using AI and Machine Learning}:  The complexity and variety of modern cyber threats require adaptable defense strategies. Future research should explore \textit{dynamic threat adaptation mechanisms}, utilizing AI and machine learning to continuously learn from new threats and update security policies accordingly. These systems should be capable of detecting abnormal patterns, adapting to novel attack techniques, and providing real-time adjustments to safeguard cloud environments.}

    \item {\textit{Proactive Threat Detection and Automated Incident Response}:  Detecting threats before they escalate is critical to reducing the impact of cyberattacks. Future research should focus on developing \textit{proactive threat detection systems} that leverage predictive analytics and AI-based anomaly detection. Coupled with this, \textit{automated incident response systems} should be further refined to ensure that identified threats are immediately neutralized, minimizing the window of vulnerability and reducing human intervention in mitigating attacks.}

    \item {\textit{Integration of Defensive, Mitigating, and Hybrid Approaches}: The future research should address  the \textit{integration of defensive, mitigating, and hybrid cybersecurity strategies}. Combining multiple approaches allows for a multi-faceted defense mechanism that can address different layers of potential threats. By synthesizing strategies from various domains (such as perimeter defense, encryption, and machine learning), researchers can build comprehensive solutions that offer robust protection against a range of security challenges.}

    \item {\textit{Explainable AI (XAI) and Transfer Learning for Enhanced Cybersecurity}:  With the increasing reliance on AI and machine learning in cybersecurity, \textit{Explainable AI (XAI)} offers an opportunity to improve the transparency and interpretability of AI-driven defense mechanisms. Future research should explore how XAI can be applied to \textit{malicious activity detection} to provide clear explanations for AI decisions. Furthermore, \textit{transfer learning} could be leveraged to adapt security solutions developed in one domain to new, unexplored threat landscapes, enhancing the flexibility and adaptability of cloud security systems.}

    \item {\textit{Collaborative Cloud Frameworks for Threat Intelligence Sharing}:  Collaborative frameworks that facilitate the \textit{sharing of threat intelligence} across organizational and geographical boundaries are crucial for strengthening global cloud security. Future research should focus on developing \textit{secure, real-time communication protocols} and infrastructure for sharing threat data, tools, and best practices among cloud providers and clients. This collaborative approach will ensure a collective defense against increasingly sophisticated cyberattacks.}

    \item {\textit{Resilient Cloud Resource Management for Future Threats}:  As cyberattacks become more sophisticated, the resilience of cloud infrastructures must be a central focus of future research. This includes developing mechanisms that allow for the \textit{secure management and allocation of cloud resources} in a way that can withstand and recover from potential breaches. Research should explore new ways to safeguard critical data and services through redundancy, fault-tolerant architectures, and adaptive resource management strategies.}
\end{itemize}

\section{Conclusions }
This paper presents a comprehensive survey and performance analysis of cloud cybersecurity countermeasure strategies for secure resource management. It introduces a novel classification of existing strategies, highlights operational challenges, and provides comparative analyses for deeper insights. An experimental study using real-world data evaluates the effectiveness of these strategies. The key findings discuss trade-offs, open research challenges, and actionable recommendations. These include developing dynamic resource allocation algorithms, implementing AI-driven systems for proactive threat detection, and designing scalable hybrid frameworks to secure diverse cloud infrastructures. These directions aim to enhance the resilience and robustness of secure cloud resource management.

\bibliographystyle{IEEEtran}
\bibliography{bibfile}

\begin{IEEEbiography}[{\includegraphics[width=1in,height=1.25in,clip,keepaspectratio]{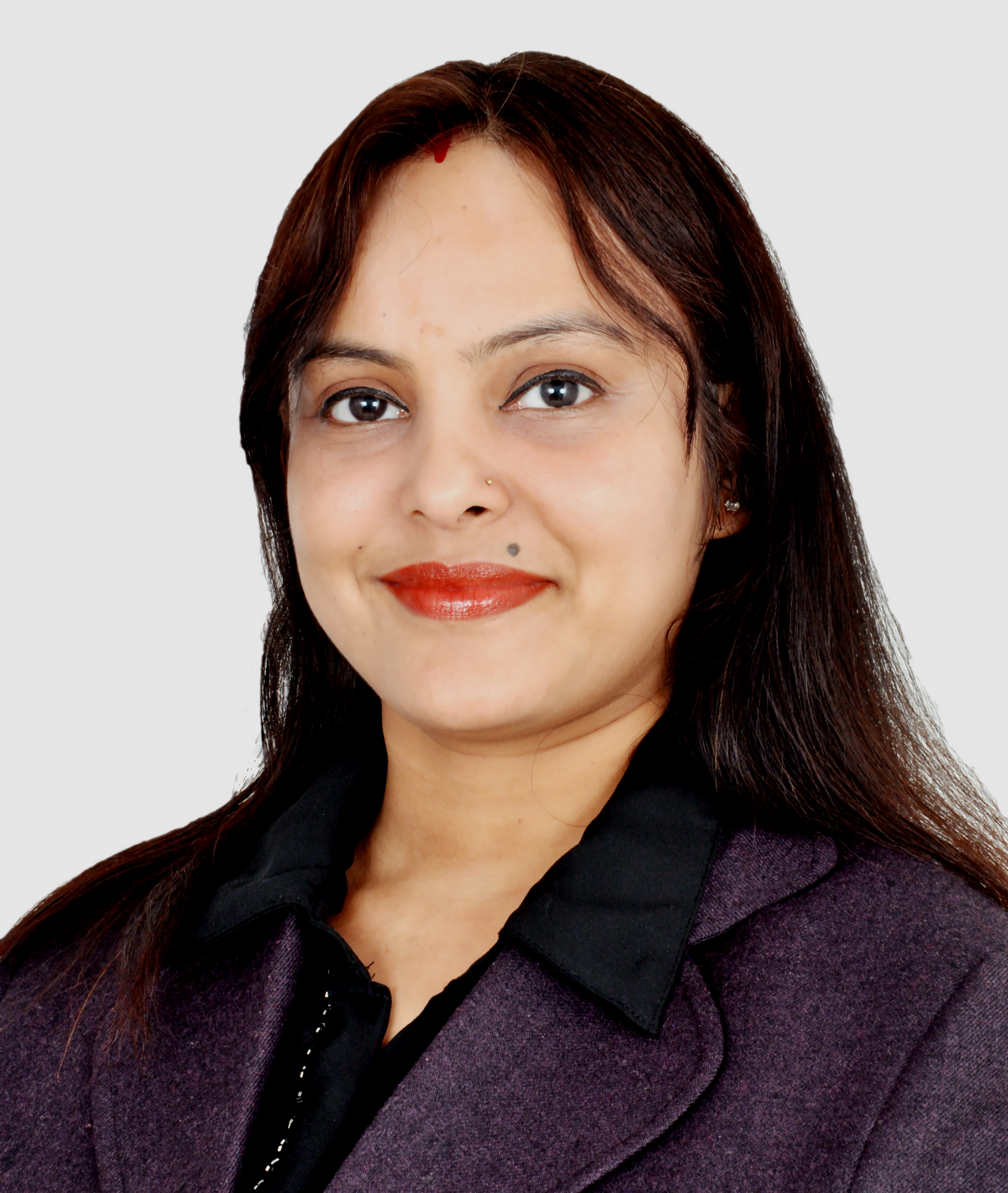}}]{Deepika Saxena} (Member, IEEE) received the Ph.D. degree in computer science from the National Institute of Technology, Kurukshetra, India, in 2022.
	
She was a Postdoctoral Fellow with the Department of Computer Science, Goethe University, Frankfurt, Germany. She is working as an Associate Professor with the Division of Information Systems, The University of Aizu, Aizuwakamatsu, Japan. She is also working as an Online Visiting Professor with the University of Economics and Human Sciences, Warsaw, Poland. Her major research interests include Neural networks, Evolutionary algorithms, resource management and security in cloud computing, Internet traffic management, and quantum machine learning, DataLakes, and dynamic caching management.

Dr. Saxena is the recipient of the prestigious IEEE TCSC Early Career Researcher Award 2024, the IEEE TCSC 2023 Outstanding Ph.D. Dissertation Award, and the EUROSIM 2023 Best Ph.D. Thesis Award. She is the recipient of the prestigious Japan Society for the Promotion of Science KAKENHI Early Career Young Scientist Research Grant FY2024. Her research paper is also published in the IEEE TRANSACTIONS ON CLOUD COMPUTING JOURNAL, received the 2022 Best Paper Award from the IEEE Computer Society Publications Board. 
\end{IEEEbiography}
\begin{IEEEbiography}[{\includegraphics[width=1in,height=1.25in,clip,keepaspectratio]{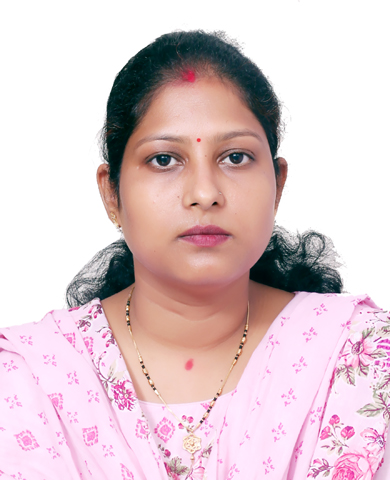}}]{Smruti Rekha Swain} received the M.Tech. degree in computer science and engineering from VSSUT, Burla, India, in 2015. She is currently pursuing the Ph.D. degree in computer science with the Department of Computer Applications, National Institute of Technology Kurukshetra, Kurukshetra, India.

Her research interests include machine learning, evolutionary algorithms, resource management, and security in cloud computing. 
\end{IEEEbiography}
 \begin{IEEEbiography}[{\includegraphics[width=1in,height=1.25in,clip,keepaspectratio]{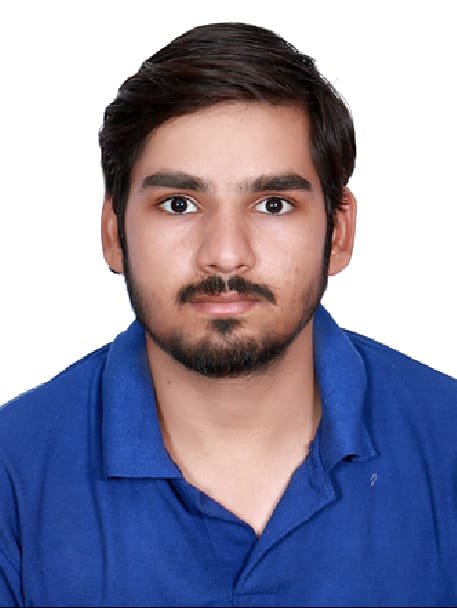}}]{Jatinder Kumar} received the B.Sc. degree in computer science from the Chaudhary Devi Lal University, Sirsa, India, in 2017, and the M.Sc. degree in computer science from the Department of Computer Science and Applications, Kurukshetra University, Thanesar, India, in 2019. He is currently pursuing the Ph.D. degree in computer science with the National Institute of Technology Kurukshetra, Kurukshetra, India.
 
 His research interests include security and privacy, cloud computing, big data, and smart grid. 
 \end{IEEEbiography}
 \begin{IEEEbiography}[{\includegraphics[width=1in,height=1.25in,clip,keepaspectratio]{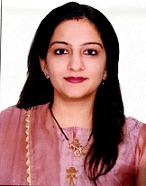}}]{Sakshi Patni} Sakshi Patni received the bachelor’s degree in computer applications from Punjab University, Chandigarh, India, in 2012, the master’s degree in computer applications from Banasthali University, Vanasthali, India, in 2015, and the Ph.D. degree in computer science from National Institute of Technology Kurukshetra, Kurukshetra, India, in 2020.
 
 From 2021 to March, 2023, she worked as an Assistant Professor with the Engineering Institutes. She is currently working as a Research Professor with the Department of Computing, Gachon University, Seongnam, South Korea. She has published various research papers in SCI, Scopus journals, and international conferences. Her main research interests include cloud computing, load balancing, resource management, federated learning, and information security.
 	
 Dr. Patni has received the Best Paper Award in IEEE International Conference ICECCS organized in Malaysia. 
 \end{IEEEbiography}
 
 \begin{IEEEbiography}[{\includegraphics[width=1in,height=1.25in,clip,keepaspectratio]{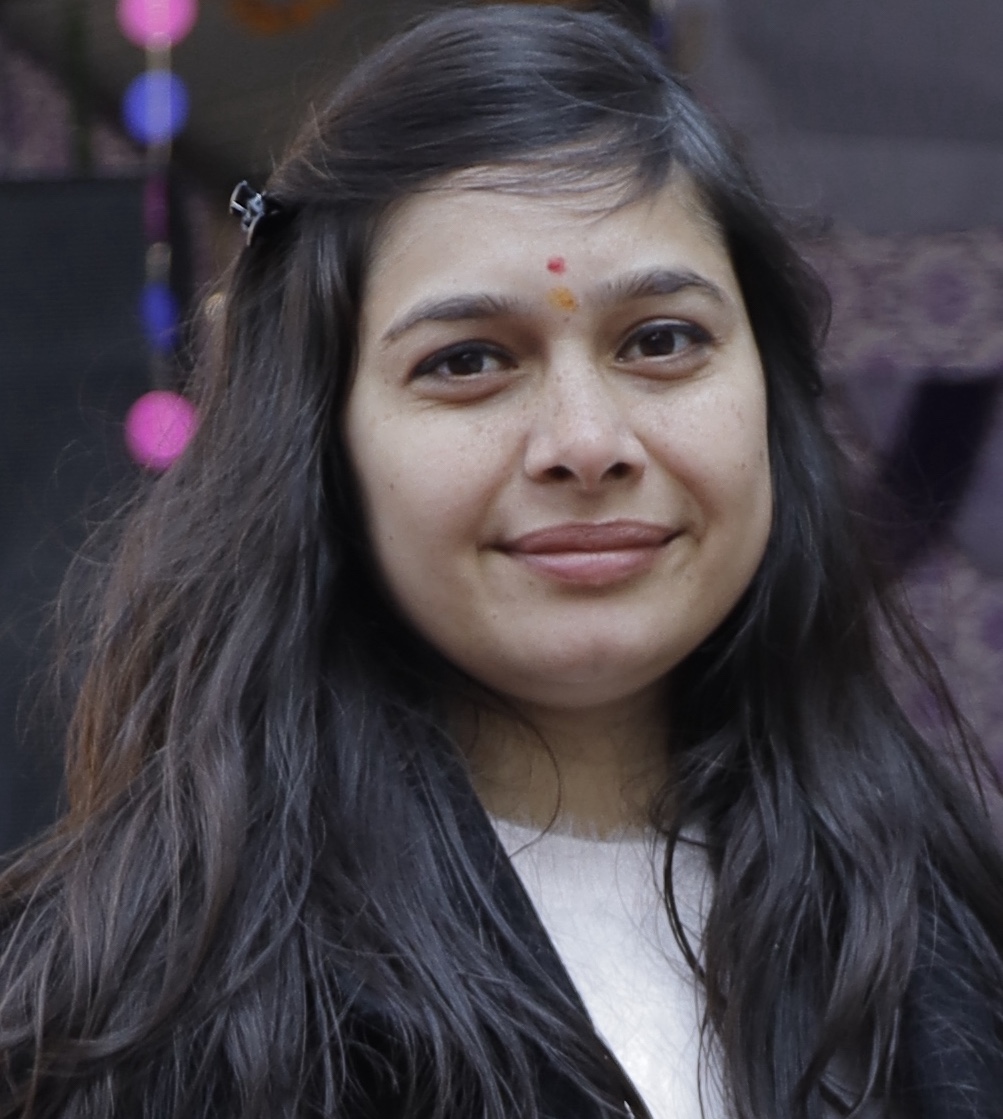}}]{Kishu Gupta} (Member, IEEE) received the Ph.D. degree in computer science from Kurukshetra University, Haryana, India, in 2023, with prestigious INSPIRE Fellowship sponsored by the Department of Science and Technology, Ministry of Science and Technology (MOST), Government of India.
 
 She is working as a Postdoctoral Research Fellow with the Cloud Computing Research Center, Department of Computer Science and Engineering, National Sun Yat-sen University, Kaohsiung, Taiwan. She is a Key Member of research projects sponsored by the MOST and National Science and Technology Council, Government of Taiwan. Her major research interests include data security and privacy, cloud computing, traffic management, federated learning, machine learning, neural networks, and quantum machine learning.
 
Dr. Gupta is also a recipient of the Gold Medal for securing first rank in overall university during M.Sc. (Computer Science). She has research findings published with top-notch venues, such as IEEE TRANSACTIONS ON AUTOMATION SCIENCE AND ENGINEERING, IEEE JOURNAL OF BIOMEDICAL AND HEALTH INFORMATICS, IEEE TRANSACTIONS ON CONSUMER ELECTRONICS, Applied Soft Computing, and Cluster Computing.
 \end{IEEEbiography}

 \begin{IEEEbiography}[{\includegraphics[width=1in,height=1.25in,clip,keepaspectratio]{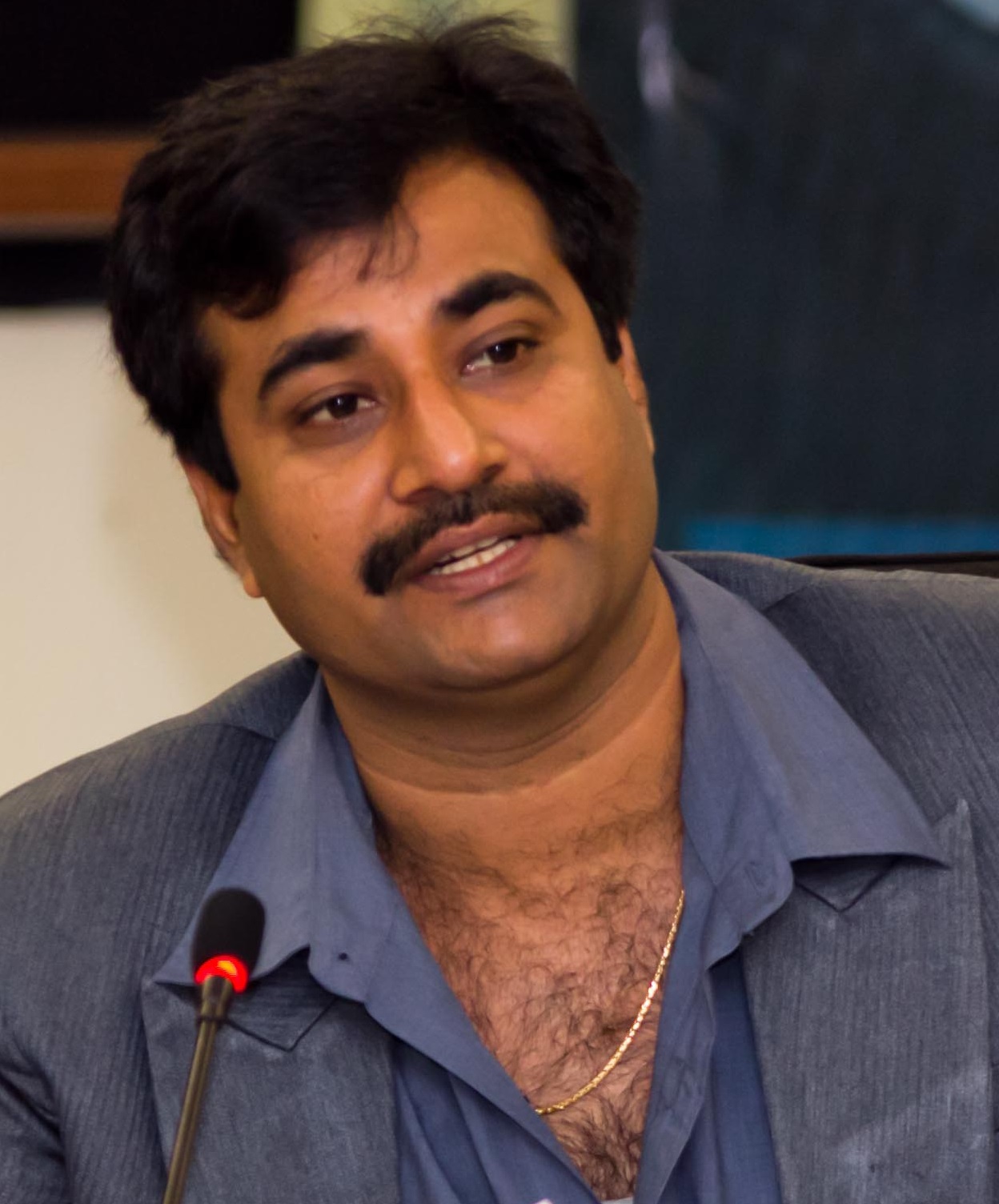}}]{Ashutosh Kumar Singh} (Senior Member, IEEE) received the Ph.D. degree in electronics engineering from the Indian Institute of Technology (BHU), Varanasi, India, in 2000.
 
 He was a Postdoctoral Fellow with the Department of Computer Science, University of Bristol, Bristol, U.K. He is working as a Professor and the Director of the Indian Institute of Information Technology Bhopal, Bhopal, India. He is also working as an Adjunct Professor with the University of Economics and Human Sciences, Warsaw, Poland. He has research and teaching experience in various Universities of the India, U.K., and Malaysia. He has published more than 400 research papers in different journals and conferences of high repute. Some of his research findings are published in top cited journals, such as IEEE TRANSACTIONS ON SERVICES COMPUTING, IEEE TRANSACTIONS ON COMPUTERS, IEEE TRANSACTIONS ON SYSTEMS, MAN, AND CYBERNETICS, IEEE TRANSACTIONS ON PARALLEL AND DISTRIBUTED SYSTEMS, IEEE TRANSACTIONS ON INDUSTRIAL INFORMATICS, IEEE TRANSACTIONS ON COMPUTERS, IEEE COMMUNICATIONS LETTERS, IEEE NETWORKING LETTERS, IEEE DESIGN AND TEST, IEEE SYSTEMS JOURNAL, IEEE WIRELESS COMMUNICATION LETTERS, IEEE TRANSACTIONS ON NETWORK AND SERVICE MANAGEMENT, IEEE TRANSACTIONS ON GREEN COMMUNICATIONS AND NETWORKING, IET Electronics Letters, FUTURE GENERATION COMPUTER SYSTEMS, Neurocomputing, Information Sciences, and Information Processing Letters. His research area includes design and testing of digital circuits, data science, cloud computing, machine learning, and security.
 
 Dr. Singh’s research paper, published in the IEEE TRANSACTIONS ON CLOUD COMPUTING JOURNAL, was honored with the 2022 Best Paper Award by the IEEE Computer Society Publications Board. 
\end{IEEEbiography}
\begin{IEEEbiography}[{\includegraphics[width=0.7\linewidth]{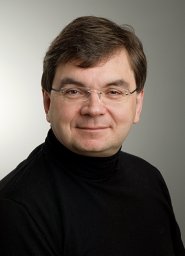}}]{Volker Lindenstruth} received the Diploma degree in physics from the Technische Universität Darmstadt, Darmstadt, Germany, in 1989, and the Ph.D. degree in physics from the GSI Helmholtz Centre for Heavy Ion Research, Darmstadt in 1993.

He then went on to work as a Postdoctoral Researcher of Computer Science for two years as a Feodor v. Lynen Fellow with LBNL, Berkeley, CA, USA. He is working as a Professor with the Department of Computer Science and the Department of Physics, Goethe University, Frankfurt, Germany. He has been the Head of the ALICE HLT Project (BMBF/CERN MoU) with the LHC of CERN, Geneva, Switzerland, and also a CERN Associate from 2006 to 2007. In 2005, he founded Certon Systems, Heidelberg, Germany. At FIAS, Frankfurt, he held the position of a Fellow since 2007 and became a Senior Fellow soon. Furthermore, the Group of High-Performance Computer Architecture, Goethe University has been in his care since 2009. His research area includes computer sciences and AI systems, energy, big data, high performance computing, and nuclear physics. 
\end{IEEEbiography}
\end{document}